\documentclass[12pt]{article}

\usepackage{graphicx}
\usepackage{amsmath,amssymb}

\usepackage{epstopdf}

\makeatletter \@addtoreset{equation}{section}

\renewcommand{\thefootnote}{\fnsymbol{footnote}}
\newcommand{\be}{\begin{equation}}
\newcommand{\ee}{\end{equation}}
\newcommand{\bear}{\begin{eqnarray}}
\newcommand{\eear}{\end{eqnarray}}
\newcommand{\ba}{\begin{array}}
\newcommand{\ea}{\end{array}}

\newcommand{\CN}{{\cal N}}

\newcommand{\tr}{{\rm tr}}

\newcommand{\bsubeq}{\begin{subequations}}
\newcommand{\esubeq}{\end{subequations}}

\def\tr{{\rm tr}}

\def\lsim{\mathrel{\rlap{\lower3pt\hbox{\hskip1pt$\sim$}}
     \raise1pt\hbox{$<$}}} %less than or approx. symbol
\def\gsim{\mathrel{\rlap{\lower3pt\hbox{\hskip1pt$\sim$}}
     \raise1pt\hbox{$>$}}}
\def\HLS1{HLS$_1$}

\DeclareFontFamily{U}{rsf}{}
\DeclareFontShape{U}{rsf}{m}{n}{
  <5> <6> rsfs5 <7> <8> <9> rsfs7 <10-> rsfs10}{}
\DeclareMathAlphabet\Scr{U}{rsf}{m}{n}
\begin{document}
\begin{titlepage}
\vfill
\begin{flushright}
{\tt\normalsize KIAS-P09008}\\
{\tt\normalsize SU-ITP-09/10}\\
%{\tt\normalsize arXiv/yymm.nnnn}\\
\end{flushright}
\vfill
\begin{center}
{\Large\bf Holographic Deuteron and \\
Nucleon-Nucleon Potential}

\vfill
Youngman Kim${}^{\spadesuit\heartsuit}$,
Sangmin Lee$^{\diamondsuit}$ and Piljin Yi$^{\spadesuit}$
%\footnote{\tt piljin@kias.re.kr}

\vskip 5mm

{\it  ${}^\spadesuit$School of Physics, Korea Institute for Advanced Study,
Seoul 130-722, Korea}
\\
{\it  ${}^\heartsuit$Asia Pacific Center for Theoretical Physics, Pohang, Gyeongbuk 790-784, Korea}
\\
{\it ${}^\diamondsuit$Department of Physics, Seoul National University,
Seoul 151-747, Korea}

\vfill
\end{center}

\begin{abstract}
\noindent
We compute the potential between a pair of nucleons in the
D4-D8 holographic QCD. In the large 't Hooft coupling
limit, $\lambda \gg 1$, the hadronic size of the baryon
is small $\sim 1/\sqrt{\lambda}M_{KK}$, and their interaction
with mesons are well approximated by a set of dimension four
and five operators. The nucleon-nucleon potential emerges
from one-boson exchange picture involving massless
pseudo-scalars and an infinite tower of spin one mesons.
We find in particular that $\rho$ meson exchanges are dominated
by a dimension five derivative coupling of tensor type,
whereas for $\omega$ mesons and axial mesons, such tensor
couplings are completely absent.
The potential is universally repulsive $\sim 1/r^2$
at short distance, and has the usual long-distance
attractive behavior $\sim -1/r^3$ along a isosinglet and
spin triplet channel. Both the large $N_c$ form and
the finite $N_c$ form are given. In the former, a shallow
classical minimum of depth $\sim 0.1M_{KK}{N_c/\lambda}$
forms at around $rM_{KK}\simeq 5.5$.

\end{abstract}

\vfill
\end{titlepage}

\tableofcontents
\renewcommand{\thefootnote}{\#\arabic{footnote}}
\setcounter{footnote}{0}

\section{Nucleon-Nucleon Potential from Holography}

The main goal of this paper is to extract the interaction
between a pair of nucleons in string theoretical framework
of holography, and consider possible bound state.
We will exclusively work with the D4-D8 model \cite{sakai-sugimoto},
which involves a large number of colors $N_c$, large 't Hooft
coupling $\lambda$, and quenching of fermions. Given many
approximations, the result should be approached with much
caution, yet we have seen often that such holographic
approaches generating realistic numbers. Ref.~\cite{Csaki:1998qr,Brower:2000rp},
for instance, gave detailed predictions on glueball spectrum of
pure QCD,
some of which were successfully compared to lattice simulation.

The D4-D8 model
has been particularly successful in encoding the spin 1
meson sector coupled with pseudo-Goldstone bosons, and
equally successful in describing baryons and the interaction
between these two sectors \cite{sakai-sugimoto,Hong:2007kx,Hata:2007mb,Hong:2007ay}.
A natural extension of the model would
be a study of the nucleon-nucleon potential.
The holographic baryon in the D4-D8 model is similar to
the Skyrmion of chiral perturbation theory. In fact, one
can view the holographic baryon as a direct uplift of the
Skyrmion in the holographic sense, which has a simple
interpretation as instanton solitons with certain
Coulombic electric hair. The size of the soliton is known
to be  $\simeq 9.6/\sqrt{\lambda}M_{KK}$ \cite{Hong:2007kx,Hata:2007mb},
where $M_{KK}$ is some natural unit in the D4-D8 model,
comparable to the lightest vector meson mass.

Given this, one can approach the problem of nucleon potential
in two different manners.
The first, which seems conceptually most natural,
is to find a ``suitable" family of two-soliton trial configuration,
emulating a pair of baryons (nucleons) taken apart from
each other, and evaluate the resulting energy. After
subtracting twice the mass of the baryon (nucleon), this
would give us a potential. However, this is much easier
said than done. The main problem with approaches like this
is that finding a ``suitable" configuration is all but
impossible for complicated solitons like this. As we will see later, the baryon-baryon
interaction scales as $N_c/\lambda$ whereas their masses
scale as $N_c\lambda$, so the interaction accounts for a very
small part of the two body energy. Unless our trial
configuration is extremely fine-tuned, energy cost due
to any slight error could easily overrun the interaction energy,
resulting in a nonsensical answer.\footnote{In fact,
approaches of this kind have been already
tried for Skyrmions with mixed results \cite{Imai:1989jh,Leese:1994hb}.}

Sometimes, however, the interaction energy
grows substantially and the physical mechanism responsible
for the interaction is easy to single out. For our solitonic
baryon this happens when the two baryons approach each
other to a distance comparable to their individual soliton
size. The leading contribution comes from the fact that
each unit soliton comes with Coulombic hair. Each baryon
has $N_c$ unit of charges with the squared electric coupling
$\sim 1/N_c\lambda$, so one finds a repulsive core interaction
of type
\begin{equation}
V_{core}\sim \frac{N_c}{\lambda}\frac{1}{M_{KK}r^2}
\end{equation}
with $r^2\simeq\lambda /M_{KK}^2$ or less.
Here $r$ denotes the mutual separation of the two baryons
and the $1/r^2$ behavior originates from the fact that the
soliton lives in approximate $R^{4+1}$.

Nevertheless, this short distance behavior
gives little insight to some common questions like how bound states,
such as deuteron and other nuclei, form. In fact, it is
not clear whether there is a realistic regime where the
precise functional behavior of the repulsive core such
as this can be
measured, since, at short distances, the asymptotic freedom
takes over and nucleons begin to see each other as collection
of partons. Also the D4-D8 model, or any other holographic
model based on gravity only, becomes somewhat dubious in the
high energy regime well beyond $M_{KK}$ because of many
non-QCD modes that begin to populate at $M_{KK}$ and higher.

If one is interested in longer distances where the interaction
is potentially attractive and where, more to the point, the
validity of the present approximation can meet real QCD,
we must consider a different approach.\footnote{The
reverse is also true. One should not be tempted to use the second
approach for the short-distance interaction, since, with much
more energy involved, the deformation of individual solitons
is inevitable.}
When the inter-baryon distance is larger that the sizes of the baryons,
the baryons can be taken to be a point-like object. In such situation,
the interactions can be all ascribed to exchange of light particles,
namely mesons. Instead of trying to understand intricate structure
of multi-solitons, one merely computes Feynman diagrams using
(cubic) interaction vertices
involving baryon currents and light mesons, such as pions $\pi$
and rho mesons $\rho$. Typical vertices that enter this computation
are
$$
\bar{\cal N}(x)\Gamma\phi(x){\cal N}(x) \,
$$
for the (pseudo-)scalar mesons, and
$$\bar{\cal N}(x)\gamma_\mu \Gamma v^\mu(x){\cal
N}(x),\qquad \bar{\cal
N}(x)\gamma^{\mu\nu}\Gamma \partial_\mu v_\nu(x){\cal N}(x)$$
for (axial-)vector mesons, with $\Gamma=1$ or $\gamma^5$.
{}From the D4-D8 holographic QCD, coupling constants for these
vertices are all precisely derivable, at least in the large $\lambda$ and
large $N_c$ limit. Then the  problem of finding nucleon-nucleon potential
becomes a matter of computing summing up tree-level Feynman diagrams
due to various meson exchanges \cite{SMR,NRS,DBS}.\footnote{See Ref.~\cite{EW88}
for a comprehensive review in the conventional approach to QCD.}

Fortunately, the basic framework for the relevant meson-nucleon
interaction has been worked out in great detail
\cite{Hong:2007kx}, where
all nucleon-meson couplings can be derived very precisely.
Some of the leading interaction strengths, such as the
leading axial coupling to pions $g_A$ and vector meson
couplings $g_{\rho{\cal NN}}$ and $g_{\omega{\cal NN}}$,
have been computed and successfully compared to experimental
data. In this note, we will take this holographic formulation
of nucleon-meson interaction and compute the nucleon-nucleon
potential from exchange of mesons. For  distances $r\gg\sqrt{\lambda}/M_{KK}$,
we find the leading potential of type
\begin{equation}
V_{exchange}\sim \frac{N_c}{\lambda}\left(\cdots\right),
\end{equation}
where the ellipsis contains terms of order $1/r$ through $1/M_{KK}^2r^3$
possibly with exponential damping factor due to vector meson
masses and also with various spin/isospin factors. Subleading
contributions start at $1/N_c\lambda$, but could be relevant in
the real QCD regime of $N_c=3$.

However,
this is not to say that the underlying
mechanism for the interaction energy is different from what one
would have obtained from the soliton approach if the latter were possible at all.
{\it Rather, the
tree diagrams involving mesons and baryons keep track of the
classical effect on one baryon by another far away, and vice versa, and
extract the classical interaction energy automatically. This
is possible because the baryons are really solitons made out of
these mesons, to begin with.}

In section 2 and 3, we give a  bare-bone review
of D4-D8 holographic QCD and the solitonic baryon thereof.
In section 4, we derive relevant meson-nucleon couplings
with emphasis on how they scale with $\lambda$ and $N_c$.
In doing so, we will learn that certain derivative
coupling of vector mesons, sometimes referred to as
the tensor coupling, can be dominant over the usual
minimal type couplings. We extract the values of these
couplings in the large $N_c$ limit.

Section 5 and 6 summarize the resulting nucleon-nucleon
potential, after suitable truncation by mass, for large $N_c$
and finite $N_c$ respectively.
It will become clear that the leading contributions come
from exchange of $\pi$, $\omega$, and axial-vector mesons
via the minimal couplings, and from exchange of $\rho$ mesons
via the tensor coupling. $\omega$
meson exchange is universally repulsive and represents a
remnant of the core repulsion we mentioned earlier,
while $\rho$ exchange off-set much of the pion exchange.
For very short distances $\sim /\sqrt{\lambda}M_{KK}$, where
the current approach become unreliable due to backreaction
of the individual solitons, the potential turns universally
repulsive as $\sim 1/r^2$ as noted above.

Section 7 gives a simplistic view of deuteron emerging from the
large $N_c$ form of the potential, and we close with a summary in
section 8.

\section{A D4-D8 Holographic QCD }

One starts with a stack of D4 branes which is compactified
on a thermal circle \cite{Witten:1998zw}, where one requires anti-periodic boundary condition
on all fermions along the circle. The purpose of having a spatial ``thermal"
circle is to give mass to the fermionic superpartners and thus break
supersymmetry. By putting $N_c$ D4 branes on a thermal circle, we obtains
pure $U(N_c)$ Yang-Mills theory in the remaining noncompact $3+1$ dimensions.
We are interested in large $N_c$ limit, so the $U(1)$ part can
be safely ignored, and we may pretend that we are studying $SU(N_c)$
theory instead. One then extrapolates the  AdS/CFT \cite{Maldacena:1997re}
to this non-conformal
case, which states that, instead of studying strongly coupled
large $N_c$ Yang-Mills theory, one may look at its dual closed string
theory. The correct dual geometry is known to be \cite{Gibbons:1987ps}
\begin{equation}
ds^2=\left(\frac{U}{R}\right)^{3/2}\left(\eta_{\mu\nu}dx^{\mu}dx^{\nu}+f(U)d\tau^2\right)
+\left(\frac{R}{U}\right)^{3/2}\left(\frac{dU^2}{f(U)}+U^2d\Omega_4^2\right) \;,
\end{equation}
with $R^3=\pi g_sN_cl_s^3$ and $f(U)=1-U_{KK}^3/U^3$.
The topology of the spacetime is $R^{3+1}\times D\times S^4$,
with the coordinate $\tau$ labeling the azimuthal angle of the disk $D$,
with $\tau=\tau+\delta\tau$ and
$\delta\tau=4\pi R^{3/2}/(3U_{KK}^{1/2})$. The circle parameterized by $\tau$
is the thermal circle.
The dilaton is
\begin{equation}
e^{-\Phi}=\frac{1}{g_s}\left(\frac{R}{U}\right)^{3/4} \:,
\end{equation}
while the antisymmetric Ramond-Ramond background field $C_{3}$
is such that $dC_3$ carries $N_c$ unit
of flux along $S^4$.

To add mesons, one introduces $N_F$ D8 branes, which share
the coordinates $x^\mu$ with the above D4 branes \cite{sakai-sugimoto} and are
transverse to the thermal circle $\tau$.
If we had not traded off the $N_c$ D4 branes in favor of the dual gravity
theory, this would have allowed massless quarks as open strings ending on
both the D4 and the D8 branes. As the D4's are replaced by the dual geometry,
however, the 4-8 open strings have to be paired up into 8-8 open strings,
the lightest of which belongs to a $U(N_F)$ gauge field, and these
are naturally identified as bi-quark mesons.
The  $U(N_F)$ gauge theory on D8 branes has the action
\begin{equation}
-\frac{4\pi^2l_s^4\mu_8}{8}\int \sqrt{-h_{8+1}}\;e^{-\Phi}\;{\rm tr} {\cal F}^2
+
\mu_8\int\,C_3 \wedge {\rm Tr}\,e^{2\pi\alpha' {\cal F} }\,,
\end{equation}
where the contraction is via the induced metric of D8 and
$\mu_p={2\pi}/{(2\pi l_s)^{p+1}} $
with  $l_s^2=\alpha'$.  The induced metric on the D8 brane is
\begin{equation}
h_{8+1}=\frac{U^{3/2}(w)}{R^{3/2}}\left(dw^2+\eta_{\mu\nu}dx^{\mu}dx^{\nu}\right)
+\frac{R^{3/2}}{U^{1/2}(w)} d\Omega_4^2\:,
\end{equation}
after we trade off the holographic (or radial) coordinate $U$ in favor of
a conformal one $w$ as\footnote{This $w$ coordinate is related to
another convenient choice of radial coordinate $z$
\begin{equation}
U^3=U_{KK}^3+U_{KK}z^2 \:  ,
\nonumber
\end{equation}
as
\begin{equation}
d(wM_{KK}) %=\frac{M_{KK}R^{3/2}}{\sqrt{U^3-U_{KK}^3}}\;dU
= \left(\frac{U_{KK}}{U}\right)^2\:\frac{dz}{U_{KK}}
=\frac{1}{(1+z^2/U_{KK}^2)^{2/3}}d(z/U_{KK}).
\nonumber
\end{equation}}
\begin{equation}
w=\int_{U_{KK}}^U{R^{3/2}dU^\prime}/{\sqrt{{U^\prime}^3-U_{KK}^3}}\:,
\end{equation}
which resides in a finite interval of length $\sim O(1/M_{KK})$ where
$
M_{KK}\equiv 3U_{KK}^{1/2}/2R^{3/2} \:.
$
Thus, the topology of the $D8$ worldvolume is $R^{3+1}\times I\times S^4$.
The nominal Yang-Mills coupling $g_{YM}^2$ is related to the other
parameters as
\begin{equation}
g_{YM}^2=2\pi g_s M_{KK} l_s \;.
\end{equation}
The low energy
parameters of this holographic theory are $M_{KK}$ and $\lambda$, which
together with $N_c$ set all the physical scales such as the QCD scale
and the pion decay constant.

In the low energy limit,  this is reduced to a
five-dimensional Yang-Mills theory with a Chern-Simons term
\begin{eqnarray}\label{five}
-\frac14\;\int_{4+1}
\; \frac{1}{e(w)^2}\sqrt{-h_{4+1}}\;{\rm tr} {\cal F}^2
+\frac{N_c}{24\pi^2}\int_{4+1}\omega_{5}({\cal A})\: ,\label{dbi}
\end{eqnarray}
where the position-dependent Yang-Mills coupling of this flavor
gauge theory is
\begin{equation}
\frac{1}{e(w)^2}=\frac{e^{-\Phi}V_{S^4}}{2\pi (2\pi l_s)^5}%= \frac{8\pi^2 R^3U(w)}{3 (2\pi l_s)^5 (2\pig_s)}
=
\frac{\lambda N_c}{108\pi^3}M_{KK}\frac{U(w)}{U_{KK}}\:,
\end{equation}
with $V_{S^4}$ the position-dependent volume of $S^4$.
The Chern-Simons coupling with $d\omega_5({\cal A})={\rm tr} {\cal F}^3$
arises because $\int_{S^4}dC_3\sim N_c$.

The usual Kaluza-Klein reduction results in an infinite number
of vector fields, whose action can be derived explicitly as
\begin{equation}
\int dx^4\,{\cal L}=\int dx^4 \sum_{n\ge 1}\tr\;\left\{{1\over 2} {\cal F}_{\mu\nu}^{(n)}
{\cal F}^{\mu\nu(n)}+m_{(n)}^2 v_\mu^{(n)} v^{\mu(n)}\right\}+\cdots\:,
\end{equation}
with ${\cal F}^{(n)}_{\mu\nu}=\partial_\mu v^{(n)}_\nu-\partial_\nu
v^{(n)}_\mu$. When we decomposed $U(N_F)$ into $SU(N_F)$ and $U(1)$,
the natural gauge generators are normalized as $\tr\, T^2=1/2$,
which explains $1/2$ in front of the kinetic
term.\footnote{In the published and all prior versions of
Ref.~\cite{Hong:2007ay}, the kinetic terms of vector and axial-vector mesons
were normalized with $1/4$ in front of the kinetic term
before the trace. % While internally
%consistent, this choice of convention is potentially misleading for
%casual readers.
With canonical normalization for (axial-)vector
mesons, the cubic couplings involving a vector or  an axial-vector meson
there should be all multiplied by $\sqrt{2}$.}
These fields can be seen as non-zero modes in
the decomposition of the gauge field, which in the (somewhat illegal but
convenient) axial gauge ${\cal A}_w=0$ is
\begin{equation}\label{ex}
{\cal A}_\mu(x;w)= i\alpha_\mu(x)\psi_{(0)}(w)+i\beta_\mu(x) +\sum_{n\ge 1}
v_\mu^{(n)}(x)\psi_{(n)}(w)\:.
\end{equation}
The eigenfunctions $\psi_{(n)}$ obey the orthonormality conditions,
\begin{equation}
\int dw\,\frac{1}{2e(w)^2}\,\psi_{(n)}(w)^*\psi_{(m)}(w)=\delta_{nm}\:,
\end{equation}
for $n,m\ge 1$. For later purpose, it is useful to introduce
\begin{equation}
\hat \psi_{(n)}(\hat w)=\sqrt{\frac{216\pi^3}{\lambda N_c}}\psi_{(n)}(w)
\end{equation}
whose form is insensitive to $\lambda N_c$. Here and in what follows, $\hat w\equiv wM_{KK}$.
Because ${\cal A}$
has a specific parity, the parity of $v_n$'s are determined by the
parity of the eigenfunctions $\psi_{(n)}(w)$ along the fifth direction.
Since the parity of any one-dimensional eigenvalue system alternates,
an alternating tower of vector and axial-vector fields emerge as the
masses $m_{(n)}$ of the KK modes increase.

To understand this zero mode part, captured in part by the
nonnormalizable eigenfunction, $\psi_{(0)}$. it is better to give up the axial
gauge and consider the Wilson line,
\begin{equation}
U(x)=e^{i\int_w {\cal A}(x,w)} \;,
\end{equation}
which, as the notation suggests, one identifies with the pion field $
U(x)=e^{2i\pi(x) /f_\pi}$.
Upon taking a singular gauge transformation back
to ${\cal A}_w=0$, one finds that it is related to $\alpha$ and $\beta$ as
\begin{eqnarray}
\alpha_\mu(x)\equiv  \{U^{-1/2},\partial_\mu U^{1/2}\} \;,\quad
2\beta_\mu(x)\equiv [U^{-1/2},\partial_\mu U^{1/2}] \;.
\end{eqnarray}
Truncating to this zero mode sector reproduces a Skyrme
Lagrangian of pions \cite{skyrme} as a dimensional reduction of the
five-dimensional Yang-Mills action,
\begin{equation}\label{Skyrme}
\int dx^4 \;\left({f_\pi^2\over 4}{\rm tr} \left(U^{-1}\partial_\mu
U\right)^2 +{1\over 32 e^2_{Skyrme}}
{\rm tr} \left[ U^{-1}\partial_\mu U, U^{-1} \partial_\nu U\right]^2\right)\:,
\end{equation}
with $ f_\pi^2=(g_{YM}^2 N_c) N_c M_{KK}^2/54\pi^4$ and
$1/e^2_{Skyrme}\simeq  {61 (g_{YM}^2 N_c)N_c}/54\pi^7$.
No other quartic terms arise, nor do we find higher order terms
in derivative, although we do recover the Wess-Zumino-Witten term
from the Chern-Simons term \cite{sakai-sugimoto}.
To compare against actual QCD, we must fix $\lambda=g_{YM}^2N_c\simeq 17$ and
$M_{KK}\simeq 0.94$ GeV
to fit  both the pion decay constant $f_\pi$ and the mass of the first vector  meson.

\section{Holographic Baryons}

The five-dimensional effective action for the $U(N_F)$ gauge field
in Eq.~(\ref{five}) admits solitons which carry a Pontryagin number
\begin{equation}
\frac{1}{8\pi^2}\int_{R^3\times I} {\rm tr} F\wedge F=k \: ,
\end{equation}
with integral $k$. We denoted by $F$ the non-Abelian part of
${\cal F}$ (and similarly later, $A$ for the non-Abelian part of ${\cal A}$).
The smallest unit  with $k=1$ carries quantum numbers of the unit
baryon.

The easiest way to see this identification
is to relate it to the Skyrmion \cite{skyrme} of chiral
perturbation theory, which is the natural object in the large $N_c$
limit \cite{'tHooft:1973jz} of QCD.
Recall that both instantons and Skyrmions are labeled by the
third homotopy group $\pi_3$ of a group manifold, which is the
integer for any semi-simple Lie group manifold $G$. For the
Skyrmion, the winding number shows up in the classification of maps
\begin{equation}
U(x):R^3\rightarrow SU(N_F=2) \;,
\end{equation}
while for the instanton it shows up as winding number at
infinity,
\begin{equation}
A(x,w\rightarrow\pm\infty)=ig_\pm(x)^\dagger dg_\pm(x) \;,
\end{equation}
with
\begin{equation}
g_-(x)^\dagger  g_+(x):R^3\rightarrow SU(N_F) \;.
\end{equation}
The relationship between the two types of the
soliton is immediate \cite{Atiyah:1989dq} once we
identify
\begin{equation}
U(x)=g_-(x)^\dagger g_+(x)\:.
\end{equation}
Therefore, the instanton soliton
in five dimensions is  the holographic image of the Skyrmions in
four dimensions. We will call it the instanton soliton.

Unlike the usual Yang-Mills theory in flat $R^4$ background,
the effective action has a position-dependent
inverse Yang-Mills coupling $1/e(w)^2$ which is a monotonically
increasing function of $|w|$. Since the Pontryagin density
contributes to the action as multiplied by $1/e(w)^2$, this tends to
position the soliton near $w=0$ and also shrink it for the same
reason. The $F^2$ energy of a trial configuration with size $\rho$
can be estimated easily in the small $\rho$ limit,\footnote{ The estimate of energy
here takes into account the spread of the instanton density
$D(x^i,w)\sim \rho^4/(r^2+w^2+\rho^2)^4$, but ignores
the deviation from the flat geometry along the four
spatial directions.}
\begin{equation}
E_{\rm Pontryagin}= \frac{\lambda N_c}{27\pi}M_{KK}\times \left(1+
 \frac16\, M_{KK}^2\rho^2+\cdots\right)\:,
\end{equation}
which clearly shows that the energy from the kinetic term increases
with $\rho$. This by itself would collapse the soliton to a
point-like one, making further analysis impossible.

A second difference comes from the presence of the additional
Chern-Simons term $\sim {\rm tr} {\cal A\wedge F\wedge F}$, whereby
the Pontryagin density $F\wedge F$ sources some of the gauge
field ${\cal A}$ minimally.
This electric charge density costs the Coulombic energy
\begin{equation}\label{C}
E_{\rm Coulomb}\simeq \frac12\times \frac{e(0)^2N_c^2}{10\pi^2\rho^2}+\cdots\:,
\end{equation}
again in the limit of $\rho M_{KK} \ll 1$. This Coulombic energy
tends to favor larger soliton size, which competes against the
shrinking force due to $E_{Pontryagin}$.

The combined energy is minimized at \cite{Hong:2007kx,Hata:2007mb,Hong:2007ay}
\begin{equation}\label{size}
\rho_{baryon}\simeq \frac{({2\cdot 3^7\cdot\pi^2/5})^{1/4}}{M_{KK}\sqrt\lambda }\:,
\end{equation}
and the  classical mass of the stabilized soliton  is
\begin{eqnarray}\label{mass}
m_B^{classical}&=&\left(E_{\rm Pontryagin}+E_{\rm Coulomb}\right)\biggr\vert_{\rm minimum} \nonumber\\
&=&\frac{\lambda N_c}{27\pi}M_{KK}
\times\left(1+\frac{\sqrt{2\cdot 3^5\cdot \pi^2/5}}{\lambda} +\cdots\right)\;.
\end{eqnarray}
As was mentioned above, the size $\rho_{baryon}$ is significantly
smaller than $\sim 1/M_{KK}$. We have a
classical soliton whose size is a lot smaller than the fundamental
scale of the effective theory.\footnote{This tendency of the baryonic
soliton shrinking to smaller size can be understood as being due to the
backreaction of vector and axial vector mesons on the conventional
Skyrmion \cite{vector-skyrmion}.}

For the sake of simplicity, and also because
the quarks in this model have no bare mass,
we will take $N_F=2$ for the rest of the note.
A unit instanton soliton in question comes with six
collective coordinates. Three correspond to the position in $R^3$, and
three correspond to the gauge angles in $SU(N_F=2)$. If the soliton is
small enough ($\rho M_{KK}\ll 1$), there exists
approximate symmetries $SO(4)=SU(2)_+\times SU(2)_-$ at $w=0$, so the total rotational symmetry
of a small solution at origin is $SU(N_F=2)\times SU(2)_+\times SU(2)_-$.
The instanton can be rotated by a conjugate $SU(2)$ action as,
\begin{equation}
F\quad\rightarrow\quad S^\dagger FS \:,
\end{equation}
with any $2\times 2$ special unitary matrices $S$ which span
${\mathbf S}^3$. Then, the quantization of the soliton is a matter
of finding eigenstates of free and nonrelativistic nonlinear
sigma-model onto ${\mathbf S}^3$ \cite{Finkelstein:1968hy,ANW}.
Under the approximate symmetry $SU(N_F=2)\times SU(2)_+\times SU(2)_-$,
the quantized instantons are in \cite{Park:2008sp}
\begin{equation}
(2s+1;2s+1;1) \;,
\end{equation}
while the quantized anti-instantons are in
\begin{equation}
(2s+1;1;2s+1) \;.
\end{equation}
Possible values for $s$ are integers and half-integers. However,
we are eventually interested in $N_c=3$, in which case spins and
isospins are naturally half-integral.  Thus we will subsequently
consider the case of $s=1/2$ states only, which are nucleons. Exciting these isospin
comes at energy cost.

\section{Nucleon-Meson Interactions from Holography \label{NNM}}

\subsection{General Formulation}

The starting point is the five-dimensional effective action of
isospin 1/2 baryons. With
\begin{equation}
\gamma^0=\left(\begin{array}{rr} 0 & -1 \\ 1 & 0\end{array}\right),\quad
\gamma^i=\left(\begin{array}{cc} 0 & \sigma_i \\ \sigma_i & 0\end{array}\right),\quad
\gamma^5=\left(\begin{array}{rr} 1 & 0 \\ 0 & -1\end{array}\right) ,
\end{equation}
we have the following five-dimensional
effective action,
\begin{eqnarray}
&&\int d^4 x dw\left[-i\bar{\cal B}\gamma^m D_m {\cal B}
-i m_{\cal B}(w)\bar{\cal B}{\cal B} +{2\pi^2\rho_{baryon}^2\over
3e^2(w)}\bar{\cal B}\gamma^{mn}F_{mn}{\cal B} \right]\nonumber \\
&-&\int d^4 x  dw {1\over 4 e^2(w)} \;{\rm tr}\, {\cal F}_{mn}{\cal F}^{mn}\,,
\label{5d}
\end{eqnarray}
where the covariant derivative is defined as $D_m=\partial_m-i(N_c {\cal A}_m^{U(1)}+{A}_m)$
with $A_m$ in the fundamental representation of $SU(N_F=2)$.
The position-dependent mass $m_{\cal B}(w)\simeq 4\pi^2 /e(w)^2\times (1+O(1/\lambda))$ is a very sharp
increasing function of $|w|$, such that in the large $N_c$ and large $\lambda$ limit,
the baryon wavefunction is effectively localized at $w=0$. This is the limit
where the above effective action is trustworthy.

The vertex $\bar{\cal B}F{\cal B}$ has the coefficient function,
about which we only know the central value precisely as
\begin{equation}
{2\pi^2\rho_{baryon}^2\over 3e^2(0)} = \frac{N_c}{\sqrt{30}}\cdot\frac{1}{M_{KK}}\;,
\end{equation}
which shows that this second interaction vertex can be actually dominant over the minimal coupling,
although it looks subleading in the derivative expansion. As it turns out,
this term is dominant for cubic vertex processes involving pions or axial
vector mesons \cite{Hong:2007ay}. How to continue this coefficient function
to $w\neq 0$ is unknown. However, for all large $\lambda N_c$ estimate of
nucleon-meson interaction terms, only this central value matters.
We chose to use the specific form above for a later convenience
but it is important to remind ourselves that the precise choice
does not matter.

To obtain interactions between nucleons and mesons,
we mode expand ${\cal B}(x^\mu,w)=B_{+}(x^\mu)f_{+}(w)+B_{-}(x^\mu)f_-(w)$
where $\gamma^5 B_{\pm}=\pm B_{\pm}$
and the profile functions $f_\pm(w)$ satisfy
\begin{eqnarray}
\partial_w f_+(w)+m_{\cal B}(w) f_+(w) &=& m_{\cal N} f_-(w)\:,\nonumber\\
-\partial_w f_-(w)+m_{\cal B}(w) f_-(w) &=& m_{\cal N} f_+(w)\:,
\end{eqnarray}
in the range $w\in[-w_{max},w_{max}]$. The 4D Dirac field for
the nucleon is then reconstructed as
\begin{equation}
{\cal N}= B_+ + B_- \:,
\end{equation}
The eigenvalue $m_{\cal N}$ is the mass of the nucleon mode ${\cal N}(x)$.
Approximating $m_{\cal B}(w)\simeq m_B^{classical}(1+(wM_{KK})^2/3+\cdots)$,
we find
\begin{equation}
m_{\cal N}\simeq m_B^{classical}+O(M_{KK})\:,
\end{equation}
so for large $\lambda$ and large $ N_c$ limit, we can take
$m_{\cal N}\simeq m_B^{classical}\simeq \lambda N_c M_{KK}/27\pi$.\
The eigenfunctions $f_\pm(w)$ are also normalized as
\begin{equation}
\int_{-w_{max}}^{w_{max}} dw\,\left|f_+(w)\right|^2 =
\int_{-w_{max}}^{w_{max}} dw\,\left|f_-(w)\right|^2 =1\:,
\end{equation}
Note that there is  a 1-1 mapping of eigenmodes
with $f_-(w)=\pm f_+(-w)$, where the sign choice is tied to
the sign choice for $m_{\cal N}$. Due to the
asymmetry under $w\to -w$, $f_+(w)$ tends to shift to
the positive $w$ side, and the opposite happens for $f_-(w)$.
In this note, we will take the convention where
$f_-(w)=f_+(-w)$. Both can be taken to be real.

Inserting this  into the action (\ref{5d}),  we find the following structure
of the four-dimensional nucleon action
\begin{equation}
\int dx^4\;{\cal L}_4 = \int dx^4\left(-i\bar {\cal N}
\gamma^\mu\partial_\mu {\cal N}-im_{\cal N}\bar {\cal N}{\cal N}+ {\cal
L}_{\rm vector} +{\cal L}_{\rm axial}\right)\:,
\end{equation}
where we have, schematically, the vector-like cubic couplings
\begin{equation}
{\cal L}_{\rm vector}=-i\bar {\cal N} \gamma^\mu \beta_\mu
 {\cal N}-\sum_{k \ge 1}g_{V}^{(k)} \bar {\cal N} \gamma^\mu  v_\mu^{(2k-1)}
 {\cal N}+\sum_{k \ge 1}g_{dV}^{(k)} \bar {\cal N} \gamma^{\mu\nu}  \partial_\mu v_\nu^{(2k-1)}
 {\cal N}\:,\label{vector-coupling}
\end{equation}
and the axial cubic couplings to axial mesons,
\begin{equation}
{\cal L}_{\rm axial}=-\frac{i g_A}{2}\bar {\cal N}  \gamma^\mu\gamma^5
\alpha_\mu {\cal N} -\sum_{k\ge 1} g_A^{(k)} \bar {\cal N} \gamma^\mu\gamma^5
v_\mu^{(2k)} {\cal N}+\sum_{k\ge 1}g_{dA}^{(k)} \bar {\cal N} \gamma^{\mu\nu}  \gamma^5\partial_\mu v_\nu^{(2k)}
 {\cal N}\:.
\end{equation}
For instance, $g_A$ is the axial coupling to pions, whose leading
cubic coupling to ${\cal N}$ appears via
\begin{eqnarray}
\alpha_\mu^{SU(2)}&=&\{\xi^{-1},\partial_\mu\xi \}^{SU(2)}={2i\over f_\pi}\partial_\mu
\pi+\cdots = {2i\over f_\pi}\partial_\mu
\pi^a\frac{\tau^a}{2}+\cdots \:.
\end{eqnarray}
We will ignore quartic couplings involving more than one spin 1 mesons.

We must recall an important detail which is suppressed in the
notation above, regarding
the differences between the isospin singlet mesons and the triplet
mesons. These two are packaged into the five-dimensional gauge field
${\cal A}$ as the trace part and the $SU(2)$ part, respectively. For instance,
the vector mesons would show up in ${\cal A}$ as
\begin{equation}
v_\mu^{(2k-1)}=\left(\begin{array}{cc}1/2 &0 \\ 0&1/2\end{array}\right)
\omega^{(k)}_\mu+\rho_\mu^{(k)a}\frac{\tau^a}{2}\:,
\end{equation}
where $\omega$'s and $\rho$'s are canonically normalized. A crucial point
is that the representation of ${\cal A}$ that appears in the baryon effective
action is different from this. Instead, the vector meson that enters the
baryon vertex has the form
\begin{equation}
v_\mu^{(2k-1)}=\left(\begin{array}{cc}N_c/2 &0 \\ 0&N_c/2\end{array}\right)
\omega^{(k)}_\mu+\rho_\mu^{(k)a}\frac{\tau^a}{2}\:,
\end{equation}
implying the isosinglet has a relative enhancement factor of $N_c$.
The second difference can be seen in the fact that only the isotriplets
appear in the  $\bar{\cal B}F{\cal B}$ vertex in five
dimensions. In any case, each and every cubic
coupling above comes in two different varieties, ones for isosinglet mesons,
such as $\eta$ and $\omega$, and those for isotriplet mesons, such as $\pi$ and $\rho$.

\subsection{Structure of the Cubic Couplings}

All the  coupling constants $g_{V,A}^{(k)}$, $ g_{dV,dA}^{(k)}$,
and $g_A$ are calculated by suitable wave-function overlap integrals
involving $f_\pm$ and $\psi_{(n)}$'s \cite{Hong:2007ay}.\footnote{A related but different
approach to these couplings was later formulated in Ref.~\cite{Hashimoto:2008zw}
which adopted the conventional methods used for Skyrmions. Since both are based
on the classical solitons quantized over the moduli space, the end results
should be equivalent.}
Let us consider the general structure. Contributions from the minimal coupling,
$\bar{\cal B}\gamma^\mu{\cal A}_\mu{\cal B}$, has the form,
\begin{equation}
A_n^\pm\equiv\int_{-w_{max}}^{w_{max}} dw\,\left|f_\pm(w)\right|^2 \psi_{(n)}(w)\:.
\end{equation}
$A_n^\pm$'s contribute to dimension four vertices, most notably
$\bar{\cal N}\gamma^\mu\rho_\mu{\cal N}$,
$\bar{\cal N}\gamma^\mu\omega_\mu{\cal N}$, and their axial vector
counterparts. They also contribute to $\bar{\cal N} \gamma^\mu\gamma^5
\partial_\mu \pi {\cal N}$, although only as a subleading contribution.

Contributions from $\bar{\cal B}F{\cal B}$ have the general forms
\begin{equation}
B_n^\pm \equiv\int_{-w_{max}}^{w_{max}}
 dw \left(\frac{2\pi^2\rho_{baryon}^2}{3e(w)^2}\right)
\, f_\mp(w)^*f_\pm(w) \psi_{(n)}(w)\:,
\end{equation}
for $\bar{\cal B}\gamma^{\mu\nu}F_{\mu\nu}{\cal B}$,
and
\begin{equation}
C_n^\pm\equiv\int_{-w_{max}}^{w_{max}}
 dw \left(\frac{2\pi^2\rho_{baryon}^2}{3e(w)^2}\right)
\,\left|f_\pm(w)\right|^2 \partial_w \psi_{(n)}(w)\:,
\end{equation}
for $\bar{\cal B}\gamma^{5\mu}F_{5\mu}{\cal B}$. The latter
two sets contribute only to the isotriplets.
$B_n^\pm$'s contribute to the derivative couplings such as
$\bar{\cal N}\gamma^{\mu\nu}\partial_\mu\rho_\nu{\cal N}$.
$C_n^\pm$'s generate the large $N_c$ leading contributions to
vertices involving isotriplet axial mesons, such as
$\bar{\cal N} \gamma^\mu\gamma^5\partial_\mu \pi {\cal N}$
and the minimal coupling to the axial vectors
$\bar{\cal N} \gamma^\mu\gamma^5a_\mu {\cal N}$.
$C$'s also contribute subleading pieces to vertices like
$\bar{\cal N}\gamma^\mu\rho_\mu{\cal N}$.

{}From these, we have the following cubic couplings for
isospin triplet mesons,
\begin{eqnarray}\label{isotriplet}
g_A^{triplet}&=&4C_0^+ +2A_0^+\,,\nonumber\\
g_V^{(k)triplet}&=&A_{2k-1}^+ + 2C_{2k-1}^+\,,\nonumber\\
g_A^{(k)triplet}&=&2C_{2k}^+ + A_{2k}^+\,,\nonumber\\
g_{dV}^{(k)triplet}&=&2B_{2k-1}^+\,,\nonumber\\
g_{dA}^{(k)triplet}&=&2B_{2k}^+\,,
\end{eqnarray}
where the mesons in these vertices are in the form such as
$\alpha^{(SU(2)}_\mu=2i/f_\pi\times(\partial_\mu\pi^a\tau^a/2)$
and $\rho^a_\mu\tau^a/2$. As we will see shortly, the second terms
for the first three lines are subleading in the $\lambda N_c$
limit, so that we actually have $g_A^{triplet}\simeq 4C_0^+ $,
$g_V^{(k)triplet}\simeq A_{2k-1}^+ $, $ g_A^{(k)triplet}\simeq 2C_{2k}^+ $.
Of these, $g_{dA}^{(k)triplet}=0$ identically, implying that
axial vectors have no derivative coupling in our approximation.

For isospin singlets, $B$ and $C$ contributions are absent so
we have
\begin{eqnarray}\label{isosinglet}
g_A^{singlet}&=&2A_0^+\,,\nonumber\\
g_V^{(k)singlet}&=&A_{2k-1}^+\,, \nonumber\\
g_A^{(k)singlet}&=& A_{2k}^+\,.
\end{eqnarray}
The mesons in these vertices are in the form such as
$\alpha^{U(1)}_\mu=2i/f_\pi\times ((N_c/2)\partial_\mu\eta')$ and
$(N_c/2)\omega_\mu$.

Remarkably, even before we go into any detail, we have a prediction
that all isospin singlet vectors and all axial-vectors have no
derivative coupling in this approximation.

\subsection{Scaling of the Cubic Couplings}

The key fact that allows us to extract large $\lambda N_c$
behavior of cubic couplings is that $m_{\cal B}(w)\sim 1/e(w)^2$
is proportional to $\lambda N_c$. Relative to the mesonic
eigenfunctions $\psi_{(n)}$, $f_\pm$ becomes more and more
concentrated at $w=0$. The two wavefunctions are slightly off-set
from the center by the amount $\sim \pm 1/(M_{KK}\lambda N_c)$
with the width of order $\sim 1/(M_{KK}\sqrt{\lambda N_c}\,)$.
This allows us to approximate $f_\pm^2$ or $f_+f_-$ by a delta
function at origin in wavefunction overlap integrals such as
$A_n^\pm$ provided that the integrand does not vanish near $w=0$.

For example, it is easy to see that
\begin{equation}
A_{2k-1}^\pm\:\rightarrow\:
\psi_{(2k-1)}(0)=\sqrt{\frac{216\pi^3}{\lambda
N_c}}\;\hat\psi_{(2k-1)}(\hat w=0)
\end{equation}
in the large $\lambda N_c$ limit. Here we also used the fact that
$\hat\psi_{(n)}(\hat w=wM_{KK})$ for $n\ge 1$ obey
\begin{equation}
\int d\hat w \;\frac{e^2(0)}{e^2(w)}\;\hat\psi_{(n)}(\hat
w)\hat\psi_{(m)}(\hat w)=\delta_{nm}
\end{equation}
and are independent of $\lambda N_c$ and of $M_{KK}$. In
particular, numerically we find $\hat\psi_{(1)}(0)\simeq 0.597$.
This number is an important ingredient of the low energy
nucleon-nucleon potential as we will find later.

$A_{2k}^\pm$'s, whose integrands vanish at $w=0$, take more care.
Using reality and the eigenmode equation for $f_\pm$,
\begin{equation}
A_{2k}^\pm= \frac{1}{m_{\cal N}}\int_{w}f_\pm(w)
\left(\mp\partial_w f_\mp(w)+m_{\cal B}(w)f_\mp(w)\right)
\psi_{(2k)}(w)\:.
\end{equation}
Since $\psi_{(2k)}$ is odd, the leading contribution arises from
the derivative piece, and we find
\begin{equation}
A_{2k}^\pm \:\rightarrow\: \pm \frac{M_{KK}}{2m_{\cal N}}
\sqrt{\frac{216\pi^3}{\lambda N_c}}\; \hat\psi_{(2k)}\,'(\hat
w=0)\:,
\end{equation}
which scales as $1/(\lambda N_c)^{3/2}$. Note that $|A_{2k-1}|\sim
(\lambda N_c)^{-1/2}\;\;\gg\; |A_{2k}|\sim (\lambda N_c)^{-3/2}$.

Evaluation of $B$'s is simpler because it involves $f_+f_-$ which
is an even function, so that
\begin{eqnarray}
B_n^\pm &=& \frac{2\pi^2\rho_{baryon}^2}{3}\int_{w}\frac{1}{e(w)^2} \,
f_\mp(w)^*f_\pm(w) \psi_{(n)}(w)\:.
\end{eqnarray}
With even $\psi$'s, this gives
\begin{eqnarray}
B_{2k-1}^\pm \:\rightarrow\: \frac{2\pi^2\rho_{baryon}^2}{3e(0)^2}
\sqrt{\frac{216\pi^3}{\lambda N_c}}\hat\psi_{(2k-1)}(0)\:,
\end{eqnarray}
whereas the odd cases vanish identically
\begin{equation}
B_{2k}^\pm = 0 \:.
\end{equation}

Evaluation of  $C_n^\pm$ proceeds similarly as $A^\pm_n$. Using
the equation of motion for $f_\pm$ again and recalling that
\begin{equation}
\frac{2\pi^2\rho_{baryon}^2}{3e(w)^2}\simeq
\frac{\rho_{baryon}^2}{6}\times m_{\cal B}(w)\:,
\end{equation}
we find
\begin{eqnarray}
C_n^\pm &=& \frac{\rho_{baryon}^2}{6}\int_{w} \, f_\pm \left(\mp
\partial_w f_\pm +m_{\cal N}f_\mp\right) \partial_w \psi_{(n)}(w)\:.
\end{eqnarray}
This gives
\begin{eqnarray}
C_{2k-1}^\pm &\rightarrow & \pm \frac{\rho_{baryon}^2}{12}M_{KK}^2
\sqrt{\frac{216\pi^3}{\lambda N_c}}\hat\psi_{(2k-1)}\,''(0)\,, \nonumber\\
C_{2k}^\pm &\rightarrow &  \frac{\rho_{baryon}^2}{6}m_{\cal N}
M_{KK} \sqrt{\frac{216\pi^3}{\lambda N_c}}\hat\psi_{(2k)}\,'(0)\,.
\end{eqnarray}
Also note that $|C_{2k}|\sim \sqrt{N_c/\lambda} \:\:\gg \:|C_{2k-1}|
\sim 1/\sqrt{\lambda^3 N_c}$.

The case of $n=0$ requires special attention since $\psi_{(0)}(w)$
is not normalizable and only its derivative, which is
normalizable, appears in the physical quantities. The conventional
choice is such that $\partial_{\hat w}\psi_{(0)}(0)=1/\pi$, which
is necessary for the familiar chiral Lagrangian to emerge from
this formulation. With this, we find
\begin{equation}
A_0^\pm\:\rightarrow\:\pm\frac{M_{KK}}{2 m_{\cal
N}}\;\frac{1}{\pi}\:,
\end{equation}
and
\begin{eqnarray}
B_0^\pm=0\:.
\end{eqnarray}
Finally, with the specific functional form $\sim 1/e(w)^2$
of the $\bar{\cal B}F{\cal B}$
coefficient, we have an analytical result,
\begin{eqnarray}
C_0^\pm = \frac{\rho_{baryon}^2
m_{\cal N}M_{KK}}{6\pi} =\frac{N_c}{\sqrt{30}}\;\frac{1}{\pi}\,.
\end{eqnarray}
These enter pion-nucleon couplings, which come with additional
factors of $1/f_\pi$ for each pion.

Note that some of the above integrals have signs sensitive to the
choice of $f_\pm$. Since $f_\pm$ are wavefunctions specific to the
chiral and the anti-chiral spinors, these $\pm$ signs for the
values of $A_{2k}^\pm$ and $C_{2k-1}^\pm$ have the net effective
of introducing a $\gamma^5$ to the vertex as a part of dimensional
reduction process, in addition to the existing Dirac matrices of
the vertices in (\ref{5d}). This is already manifest in how these
coefficients contributes to the cubic couplings in
Eq.~(\ref{isotriplet}),(\ref{isosinglet}).

\subsection{Pseudo-scalar Mesons: $\pi$ and $\eta'$ }

Starting with
\begin{equation}
-\frac{i g_A}{2}\bar {\cal N}  \gamma^\mu\gamma^5
\alpha_\mu {\cal N}\,,
\end{equation}
we restore the isotriplet and the isosinglet mesons and find
\begin{equation}
\frac{g_A^{triplet}}{2f_\pi}\bar {\cal N}  \gamma^\mu\gamma^5
\partial_\mu (\pi^a\tau^a){\cal N}+
\frac{g_A^{singlet} N_c}{2f_\pi}\bar {\cal N}  \gamma^\mu\gamma^5 \partial_\mu\eta'{\cal N}.
\end{equation}
 Since we will be considering
$N_f=2$, the distinction between $\eta$ and $\eta'$ becomes a bit
ambiguous. Here $\eta'$ denotes the trace part of the
pseudo-scalar, regardless of the number of flavors. In turn, this
is equivalent to
\begin{equation}
-\left(\frac{g_A^{triplet}}{2f_\pi}\times 2m_{\cal N}\right)\bar {\cal N} \gamma^5
(\pi^a\tau^a){\cal N}
- \left(\frac{g_A^{singlet} N_c}{2f_\pi}\times 2m_{\cal N}\right)\bar {\cal N} \gamma^5 \eta'{\cal N}.
\end{equation}
%
%To compute the meson exchange potentials in the next section, %
%it is convenient to introduce the following couplings,

\subsection{Vector Mesons: $\rho$ and $\omega$}

We will denote the isotriplet vectors by $\rho^{(k)}$ and singlets
by $\omega^{(k)}$, upon which
\begin{equation}
-\sum_{k \ge 1}g_{V}^{(k)} \bar {\cal N} \gamma^\mu  v_\mu^{(2k-1)}
 {\cal N}+\sum_{k \ge 1}g_{dV}^{(k)} \bar {\cal N} \gamma^{\mu\nu}  \partial_\mu v_\nu^{(2k-1)}
 {\cal N}
\end{equation}
separates to
\begin{eqnarray}
-\sum_{k \ge 1}\left(\frac{g_{V}^{(k)triplet}}{2} \right)\bar {\cal N}
\gamma^\mu  \rho^{(k)a}_\mu\tau^a
 {\cal N}+\sum_{k \ge 1}\left(\frac{g_{dV}^{(k)triplet}}{2}\right)
 \bar {\cal N} \gamma^{\mu\nu}  \partial_\mu \rho_\nu^{(k)a}\tau^a
 {\cal N}
\end{eqnarray}
and
\begin{eqnarray}
-\sum_{k \ge 1}\left(\frac{ N_cg_{V}^{(k)singlet}}{2}\right) \bar {\cal N} \gamma^\mu  \omega_\mu^{(k)}
 {\cal N}
\end{eqnarray}
since the singlet does not have the derivative coupling in this
approximation.

\subsection{Axial Vector Mesons: $a$ and $f$}

Similarly, the axial vector mesons couplings
\begin{equation}
 -\sum_{k\ge 1} g_A^{(k)} \bar {\cal N} \gamma^\mu\gamma^5
v_\mu^{(2k)} {\cal N}+\sum_{k\ge 1}g_{dA}^{(k)}
\bar {\cal N} \gamma^{\mu\nu}  \gamma^5\partial_\mu v_\nu^{(2k)}
 {\cal N}
 \end{equation}
can be written as
\begin{eqnarray}
-\sum_{k\ge 1}\left(\frac{g_{A}^{(k)triplet}}{2} \right)\bar {\cal N} \gamma^\mu  \gamma^5a^{(k)a}_\mu\tau^a
 {\cal N}
-\sum_{k\ge 1}\left(\frac{ N_cg_{A}^{(k)singlet}}{2}\right) \bar {\cal N} \gamma^\mu \gamma^5 f_\mu^{(k)}
 {\cal N}
\end{eqnarray}
since no derivative coupling exists for axial vectors in this
approximation.

\section{Large $N_c$ Nucleon-Nucleon Potential}

Phenomenologically the nucleon-nucleon (NN) potential is well
described by one boson exchange models. The long-range part of the
NN potential is mostly due to the pion exchange, while the
short-range repulsion is governed by the vector meson exchange.
The scalar meson exchange is responsible for the
intermediate-range of the potential. The interaction Lagrangians
for boson-nucleon couplings are, for pseudoscalar mesons: \bear
{\cal L}_P=-g_{\varphi {\cal NN}}\bar{\cal N}(x)\gamma_5\varphi(x){\cal N}(x) \, ,
\eear
 and for vector mesons:
\bear ~{\cal L}_V= -g_{v{\cal NN}}\bar{\cal N}(x)\gamma^\mu v_\mu(x){\cal
N}(x) + \frac{\tilde{g}_{v{\cal NN}}}{2m_{\cal N}}\bar{\cal
N}(x)\gamma^{\mu\nu}\partial_\mu v_\nu(x){\cal N}(x)\, , \eear
where $m_{\cal N}$ is the nucleon mass. For the D4-D8
holographic model, we saw that the derivative coupling is
absent for the isospin singlet vectors such as $\omega$.\footnote{
Note that, empirically, $\tilde g/g=3.7-6.1$ for the $\rho$-meson
(see for example \cite{SMR}), while for $\omega$-mesons the ratio
is close to zero, for instance $\tilde g/g=0.1\pm 0.2$~\cite{EW88}.}
The same is true of axial vectors, so we have only~\cite{DBS}
\bear
{\cal L}_A=- g_{a{\cal NN}}\bar{\cal N}(x)\gamma^\mu\gamma_5 a_\mu(x){\cal
N}(x) \, .
\eear
Note that we now use the convention for isovector bosons as
$\varphi=\vec\tau\cdot\vec\varphi$, $v=\vec\tau\cdot\vec v$, and
$a=\vec\tau\cdot\vec a$.

It is useful to compare our convention
to that of Ericson and Weise~\cite{EW88}, which is our primary reference on
one boson exchange potential. The Dirac matrices we used are
such that $i\gamma^\mu=\gamma^\mu_{\rm Ericson-Weise}$, which
brings us to the same convention for the nucleon field and its
free Lagrangian. In addition, we have reversed the overall sign
of the couplings from theirs as $g_{\varphi\cal NN}=-g_P$,
$g_{\rho\cal NN}=-g_V$, and $\tilde{g}_{\rho\cal NN}=-g_T$,
which is a matter of a common sign convention on meson fields.
We have no scalar field, so do not have the counterpart of their $g_S$.

The leading large $N_c$ and large $\lambda$ scaling is such that,
for pseudo-scalars ($\varphi=\pi,\eta'$)
\begin{eqnarray}
\frac{g_{\pi\CN\CN}}{2m_\CN} M_{KK} &=& \frac{g_A^{triplet}}{2f_\pi}M_{KK}
\simeq \frac{2\cdot 3\cdot\pi}{\sqrt{5}} \times \sqrt{\frac{N_c}{\lambda}},
%\; \simeq \; {8.43} \sqrt{\frac{N_c}{\lambda}},
\nonumber \\
\frac{g_{\eta' \CN\CN}}{2m_\CN} M_{KK} &=& \frac{N_c g_A^{singlet}}{2f_\pi}M_{KK}
\simeq \sqrt{\frac{3^9}{2}} \pi^2 \times \frac{1}{\lambda N_c} \sqrt{\frac{N_c}{\lambda}},
%\; \simeq \; \frac{979}{\lambda N_c } \sqrt{\frac{N_c}{\lambda}} .
\end{eqnarray}
for vectors ($v=\rho^{(k)},\omega^{(k)}$)
\begin{eqnarray}
g_{\rho^{(k)}\CN\CN} &=& \frac{g_V^{(k)triplet}}{2}
\simeq \sqrt{2\cdot3^3\cdot\pi^3}\:  \hat{\psi}_{(2k-1)}(0)
\times \frac{1}{N_c} \sqrt{\frac{N_c}{\lambda}} ,
\nonumber \\
g_{\omega^{(k)}\CN\CN} &=& \frac{N_c g_V^{(k)singlet}}{2}
\simeq  \sqrt{2\cdot3^3\cdot\pi^3}\: \hat{\psi}_{(2k-1)}(0) \times \sqrt{\frac{N_c}{\lambda}},
\nonumber \\
\frac{\tilde{g}_{\rho^{(k)}\CN\CN}}{2m_\CN}M_{KK} &=&
\frac{g_{dV}^{(k)triplet}M_{KK}}{2} \simeq
 \sqrt{\frac{2^2 \cdot3^2\cdot\pi^3}{5}}\:  \hat{\psi}_{(2k-1)}(0) \times
\sqrt{\frac{N_c}{\lambda}} ,
\end{eqnarray}
and for axial vectors ($a=a^{(k)},f^{(k)}$),
\begin{eqnarray}
g_{a^{(k)}\CN\CN} &\equiv& \frac{g_A^{(k)triplet}}{2}
\simeq \sqrt{\frac{2^2\cdot 3^2\cdot\pi^3}{5}}\:  \hat{\psi}_{(2k)}\,'(0)
\times \sqrt{\frac{N_c}{\lambda}} ,
\nonumber \\
g_{f^{(k)}\CN\CN} &\equiv& \frac{N_c g_A^{(k)singlet}}{2}
\simeq \sqrt{\frac{3^9 \cdot\pi^5}{2}} \: \hat{\psi}_{(2k)}\,'(0)
\times \frac{1}{\lambda N_c} \sqrt{\frac{N_c}{\lambda}} .
\end{eqnarray}
Note that $g_{\rho\cal NN}$ and $\tilde g_{\rho\cal NN}$ we have
derived from the D4-D8 model are of the same sign, which is consistent
with experimental results.

\subsection{Holographic Nucleon-Nucleon Potential }

In general, the one-boson exchange nucleon-nucleon
potential can be written as
\begin{equation}
V_\pi+V_{\eta'}+\sum_{k=1}^\infty V_{\rho^{(k)}}+\sum_{k=1}^\infty V_{\omega^{(k)}}
+\sum_{k=1}^\infty V_{a^{(k)}}+\sum_{k=1}^\infty V_{f^{(k)}}.
\end{equation}
We now borrow results on one-boson exchange potentials
from Ref.~\cite{NRS, EW88} for various mesons, and truncate
to the leading contributions in $1/N_c$ and in $1/\lambda$.
(For more complete forms of one boson exchange potential,
we refer to Appendix 10 of Ref.~\cite{EW88}.)
In doing so, we find that not all terms  in the above
expansion contribute at the leading order. The leading
contributions arise from the following four classes of
couplings
\begin{eqnarray}
\frac{g_{\pi{\cal NN}}M_{KK}}{2m_{\cal N}}
\sim
g_{\omega^{(k)}\cal NN}
\sim
\frac{\tilde g_{\rho^{(k)}\cal NN}M_{KK}}{2m_{\cal N}}
\sim
g_{a^{(k)}\cal NN}
\sim \sqrt{\frac{N_c}{\lambda}}\,,
\end{eqnarray}
whereas $g_{\eta'\cal NN}$ is further suppressed by $1/\lambda N_c$
and $g_{\rho^{(k)}\cal NN}$ by $1/N_c$.

For instance, the one pion exchange potential
(OPEP) would be
\bear V_{\pi}\label{pionfull}
=\left(\frac{g_{\pi\cal NN}}{2m_{\cal
N}}\right)^2\frac{m_\pi^3}{12\pi}[y_0(m_\pi r)\vec\sigma_1\cdot\vec\sigma_2
+y_2(m_\pi r) S_{12}]\vec\tau_1\cdot\vec\tau_2\, , \eear where
$S_{12}=3(\vec\sigma\cdot\hat r)(\sigma_2\cdot\hat r)
-\vec\sigma_1\cdot\vec\sigma_2$, and \bear
y_0(x)=\frac{e^{-x}}{x},~
y_2(x)=\left(1+\frac{3}{x}+\frac{3}{x^2}\right)\frac{e^{-x}}{x}\, .
\eear
However, since we are working in the D4-D8 model where %In the chiral limit,
$m_\pi=0$, the OPEP simplifies to
\bear\label{pion0}
V_{\pi}^{holographic}=\frac{1}{4\pi}\left(\frac{g_{\pi{\cal NN}}M_{KK}}{2m_{\cal
N}}\right)^2\frac{1}{M_{KK}^2r^3}\;S_{12}\;\vec\tau_1\cdot\vec\tau_2\, .
\eear
%\bear
%V_{(\eta')}=\frac{1}{4\pi}\left(\frac{g_{\eta'{\cal NN}}}{2m_{\cal
%N}}\right)^2\frac{1}{r^3}\;S_{12}\; .
%\eear
For the isospin singlet vector meson, namely $\omega^{(k)}$-mesons,
the derivative coupling is absent and the leading large $\lambda N_c$
contribution is very simple,
\bear V_{\omega^{(k)}}^{holographic}&= &\frac{1}{4\pi}\;
\left(g_{\omega^{(k)}\cal NN}\right)^2\;m_{\omega^{(k)}}\; y_0(m_{\omega^{(k)}}
r). %\no
%&+& \frac{1}{4\pi} \left(\frac{g_{\omega\cal NN}
%+\tilde g_{\omega\cal NN}}{2m_{\cal N}}\right)^2\frac{m_\omega^3}{3} [ 2y_0(m_\omega r)\vec\sigma_1\cdot\vec \sigma_2 -y_2(m_\omega r)
%S_{12}(\hat {r}) ]\, .
\eear
For $\rho^{(k)}$ which are the isospin triplet vector mesons,
the derivative coupling is dominant over the minimal coupling.
This also simplifies the potential quite a bit as
%the potentials are multiplied by the isospin operator, $\vec\tau_1\cdot\vec\tau_2$,
\bear
&&V_{\rho^{(k)}}^{holographic}\simeq\nonumber\\\nonumber\\
%&\simeq &\frac{1}{4\pi}g_{\rho^{(k)}\cal NN}^2m_{\rho^{(k)}} y_0(m_{\rho^{(k)}}r)\no
&& \frac{1}{4\pi} \left(\frac{\tilde g_{\rho^{(k)}\cal NN}M_{KK}}{2m_{\cal N}}\right)^2
\frac{m_{\rho^{(k)}}^3}{3M_{KK}^2} [ 2y_0(m_{\rho^{(k)}} r)
\vec\sigma_1\cdot\vec \sigma_2 -y_2(m_{\rho^{(k)}} r)
S_{12}(\hat {r}) ]\:\vec\tau_1\cdot\vec\tau_2\, .
\eear
The contribution to $V_{\rho^{(k)}}$ due to the minimal coupling
$g_{\rho\cal NN}$ are suppressed by additional $1/N_c$.

The potential from exchange of isospin singlet axial vectors $f^{(k)}$
is suppressed by additional $1/(\lambda N_c)^2$ while triplet axial-vector
mesons ${a^{(k)}}$ contributes~\cite{DBS}
\bear
&&V_{a^{(k)}}^{holographic}\simeq\nonumber\\\nonumber\\
&& \frac{1}{4\pi}\,\left({g_{a^{(k)}\cal NN}}\right)^2
\;\frac{m_{a^{(k)}}}{3} \;[ -2y_0(m_{a^{(k)}}
r)\vec\sigma_1\cdot\vec \sigma_2 +y_2(m_{a^{(k)}} r) S_{12}(\hat
{r})]\:\vec\tau_1\cdot\vec\tau_2 \, .
\eear
Finally note that the meson masses are all of order $M_{KK}$
and $m_{\rho^{(k)}}=m_{\omega^{(k)} } < m_{a^{(k)}}$. The vector
masses and the axial vector masses alternate as $k$ increases.

\subsection{Behavior at $r\sim 1/M_{KK}\gg 1/\sqrt{\lambda}M_{KK}$ }

When the distance in question is longer than $1/M_{KK}$, it suffices to
consider contributions from light mesons only,
\begin{equation}
V_{(p)}\equiv  V_\pi^{holographic}
+\sum_{k=1}^{p}\left( V_{\rho^{(k)}}^{holographic}+
V_{\omega^{(k)}}^{holographic}+V_{a^{(k)}}^{holographic}\right),
\end{equation}
where the level $p$ is determined by the short distance scale,
down to which we are interested. For instance, if we are
interested in distance down to $1/(3M_{KK})$, $p=10$ would suffice.

More generally, with large but finite $\lambda$, the smallest
distance where one can still trust this one-boson exchange
potential is when the distance is comparable to the solitonic
size of the nucleon at $\sim 1/\sqrt{\lambda}M_{KK}$. Around
this scale, the current set-up, where one implicitly assumes each
of the unit baryon to be intact, breaks down and one must begin
to consider backreactions systematically. Thus, although the
sum can formally extend to $p=\infty$, it is in practice more
sensible to cut it off at $p\sim \sqrt{\lambda/10}$, after taking
into accounts various order one factors.

The relevant (large $\lambda N_c$) pion coupling is
\begin{equation}
\frac{g_{\pi\CN\CN}}{2m_\CN} M_{KK} \simeq {8.43} \sqrt{\frac{N_c}{\lambda}},
\end{equation}
while for (axial-)vector mesons we parameterize the relevant coupling as
\begin{eqnarray}
%g_{\rho \CN\CN} \simeq \frac{17.3}{N_c} \sqrt{\frac{N_c}{\lambda}},
%\;\;\;\;\;
g_{\omega^{(k)} \CN\CN} \simeq %17.3
\xi_k \; \sqrt{\frac{N_c}{\lambda}} ,
\;\;\;\;\;
\frac{\tilde{g}_{\rho^{(k)} \CN\CN}}{2m_\CN}M_{KK} \simeq %6.31
\zeta_k\;\sqrt{\frac{N_c}{\lambda}} ,
\;\;\;\;\;
g_{a^{(k)}\CN\CN} \simeq \chi_k \;\sqrt{\frac{N_c}{\lambda}}.
%\;\;\;\;\;
%g_{f^{(k)}\CN\CN} \simeq ???
\end{eqnarray}
Coefficients, $\xi_k$, $\zeta_k$, $\chi_k$, are determined by $\psi_{(2k-1)}(0)$
and $\psi_{(2k)}'(0)$, we list these values in the following table 1, together with
the masses (in unit of $M_{KK}$) of the vector and the axial vector mesons.

\begin{table}[htb]
\begin{tabular}{c||c|c|c|c||c|c|c}
$\quad k\quad $ & $m_{\omega^{(k)}}=m_{\rho^{(k)}}$& $\hat\psi_{(2k-1)}(0)$
& $\quad\xi_k\quad$ & $\quad \zeta_k \quad$
& $\quad m_{a^{(k)}}\quad $ &  $\hat\psi'_{(2k)}(0)$ & $\quad\chi_k\quad$  \\
\hline
1 & 0.818 &0.5973 & 24.44 & 8.925 & 1.25 &0.629 & 9.40 \\
\hline
2 & 1.69  &0.5450 & 22.30 & 8.143 & 2.13 &1.10  & 16.4 \\
\hline
3 & 2.57  &0.5328 & 21.81 & 7.961 & 3.00 &1.56  & 23.3\\
\hline
4 & 3.44  &0.5288 & 21.64 & 7.901 & 3.87 &2.02  & 30.1\\
\hline
5 & 4.30 & 0.5270 & 21.57 & 7.874 & 4.73 & 2.47 & 36.9\\
\hline
6 & 5.17 & 0.5261 & 21.52 & 7.860 & 5.59 & 2.93 & 43.8 \\
\hline
7 & 6.03 & 0.5255 & 21.50 & 7.852 & 6.46 & 3.38 & 50.5\\
\hline
8 & 6.89 & 0.5251 & 21.48 & 7.846 & 7.32 & 3.83 & 57.3\\
\hline
9 & 7.75 & 0.5249 & 21.48 & 7.843 & 8.19 & 4.29 & 64.1\\
\hline
10 & 8.62 & 0.5247 & 21.47 & 7.840 & 9.05 & 4.74 & 70.9
\end{tabular}
\caption{\small
Numerical results for masses and coupling constants for spin one mesons
interacting with nucleons. }
\label{table1}
\end{table}

In figure 1, we display the shape of the large $N_c$ potential
with $p=10$ for the iso-singlet sector with total angular
momentum one and total spin one. By superselection rules, the spatial
angular momentum is a mixture of 0 and 2, and effectively we
have
\begin{equation}
S_{12}=2,\quad \vec \tau_1\cdot\vec \tau_2=-3,\quad \vec \sigma_1\cdot\vec\sigma_2=1.
\end{equation}
This is the only channel which is
attractive at long distance. All other channels are repulsive.
See section 7
for more discussion. The minimum
of the potential is located around $5.5/M_{KK}$ which is a little
larger than one fermi if we adopt $M_{KK}\simeq 0.94 GeV$.
Toward $r=0$, the potential becomes repulsive very quickly,
and this is consistent with the expected short distance behavior
we will see in next subsection.

\begin{figure}[t]
\begin{center}
\vskip -1cm
\scalebox{1}[1]{\includegraphics{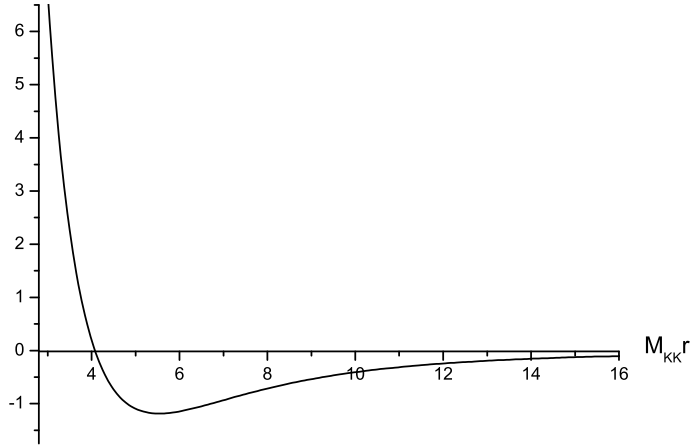}}
\par
\end{center}
\caption{\small
A plot of large $N_c$ nucleon-nucleon potential, truncated at $p=10$,
along its most attractive channel, namely isospin singlet, $\sigma$-spin
triplet, and even spatial angular momentum with $S_{12}=2$. The
horizontal axis is for the distance, $rM_{KK}$, while the potential
energy along the vertical axis is in unit of $M_{KK}N_c/4\pi\lambda$.} \label{NN10}
\end{figure}

If we continue past toward small $r$
beyond the region of validity (set by the integer $p$), the potential actually
turns attractive again very sharply. However this is an artifact of
cutting off the series at finite number of KK modes, and should not be taken
seriously. Indeed, this unphysical turnaround can be seen to occur
right below $1/3M_{KK}$ where we expect the $p=10$ formula to become
untrustworthy, at least for arbitrarily large $\lambda$.
For finite $\lambda$, however, it turns out that there is a very simple
remedy of this problem. The unphysical turnaround turns out to be
a combined effect of the truncation and certain finite $\lambda$
correction that we ignored in section 4. By choosing an optimal
value of $p$ in accordance with $\lambda$, one can easily restore
physical sensible short-distance behavior as we explain in next subsection.

%\bear
%&&V_v^{(\rho)} (r)=\frac{g_V^2}{4\pi}m_v y_0(m_v r)\no
%&&~~~~~+\frac{g_V^2}{48\pi}m_v\left(1+\frac{g_T}{g_V}\right)^2\frac{m_v^2}{m_{\cal
%N}^2} [ 2y_0(m_v r)\vec\sigma_1\cdot\vec \sigma_2 -y_2(m_v r)
%S_{12}(\hat {r}) ]\vec\tau_1\cdot\vec\tau_2\, .
%\eear
%Note that
%\bear \vec\sigma_1\cdot\vec
%\sigma_2\vec\tau_1\cdot\vec\tau_2=4[S(S+1)-\frac{3}{2}][I(I+1)-\frac{3}{2}]\,
%, \eear where $S$ and $I$ are total spin and isospin,
%respectively. Some typical values for the couplings are
%$g_P^2/4\pi\simeq 14$ and $g_V^2/4\pi\simeq 0.6$ for the $\rho$-meson.

\subsection{Coulomb Repulsion at Short Distance
and Finite $\lambda $ Corrections to the Large $N_c$ Potential}

When the distance between the pair of nucleon is much smaller
than $1/M_{KK}$ and comparable to $1/\sqrt{\lambda}M_{KK}$,
the above expressions must be summed over all mesons. When $\lambda$
is sufficiently large, however, it is clear where the leading
contribution comes from. The holographic picture of the solitonic
baryon involves an instanton soliton with a unit Pontryagin number
dressed with Abelian electric charge. When the net soliton
configuration is smaller than the curvature scale of the
background holographic geometry, $1/M_{KK}$, the instanton
part of the soliton will behave like that of ordinary
instanton on $R^4$ with scale invariance.

This implies that the leading potential energy beyond the rest masses of the
two cores should come from the five-dimensional electrostatic
energy associated with the Abelian electric charge. Roughly
each soliton has $N_c$ unit of electric charges and the five
dimensional electric coupling scales as $1/\sqrt{\lambda N_c}$,
and this gives repulsive potential
\begin{equation}
\sim \frac{N_c}{\lambda}\frac{1}{M_{KK} r^2}\,.
\end{equation}
Details of this potential are, however, more complicated. The
electric charge density is basically the same as the Pontryagin
density, so the precise form of the two-instanton solution enters
the potential. In particular the relative spatial/gauge
orientation of the two-instanton configuration must enter
the potential, predicting a particular spin/isospin-dependence.

Clearly, the precise and quantitative structure of
the short-distance potential cannot be
captured by the our one-boson exchange potential since
the underlying formulation for the latter ignores the core
shape of the soliton other than its spin/isospin structures,
whereas in the short-distance $\sim 1/\sqrt{\lambda}M_{KK}$ the
potential energy is of order $N_c$ and is comparable to the
electric part of the soliton energy.
In order to compute the precise structure of this short distance behavior, one
should at least start from the full two-instanton solution, available
in the literature either via ADHM construction or in the
form of Jackiw-Nohl-Rebbi (JNR) ansatz \cite{Jackiw:1976fs}.\footnote{While
our work was in progress, there appeared two related papers \cite{Kim-Zahed,Hashimoto:2009ys}
that share some common goal with our work. The latter in particular
worked out a precise short-distance form of the potential using ADHM
construction of two-instanton.
%We refer readers to Hashimoto et.al \cite{Hashimoto:2009ys}
%for details of short distance
%nucleon-nucleon interaction.
}
Unfortunately, however, this approach is difficult to
extend beyond very short distance, since the analog of
AHDM or JNR is not available in a curved background.
%Nor can we use the usual conformal properties of instanton, since the additional Coulombic
%hair (which is nothing but the baryon number, elevated to a gauge charge
%in the holographic setting) destroys it.

Independent of this, as a self-consistency check, we wish to understand
how the sum over the KK tower of mesons end up producing $1/r^2$
behavior at short distance. The leading short-distance power from
individual meson exchange is $1/r^3$. Since KK modes sum over such
powers (after taking into account the coefficients carefully) cannot
make a $1/r^2$ form, somehow $1/r^3$ terms must cancel in the full
summation over mesons. For instance, pions contributes $\simeq 71N_c/4\pi\lambda$
to the coefficient, whereas the $\rho$ and the first $a$ meson
contribute $\simeq -80N_c/4\pi\lambda$ and $\simeq 57N_c/4\pi\lambda$,
respectively. Continuing this fashion, one can see that the
pion contribution is gradually eaten away by the alternating
contributions from the pairs ($\rho^{(k)}$, $a^{(k)}$). However,
the sum up to $p=10$ can be seen to weaken $\sim 1/r^3$ from the
pion exchange only by a factor of half, which is not enough for the
anticipated cancellation. In numerical plot with $p=10$, shown
in figure 2, this manifests as an unphysical turnaround at
$rM_{KK}\simeq 3/10$.

\begin{figure}[t]
\begin{center}
\scalebox{1}[1]{\includegraphics{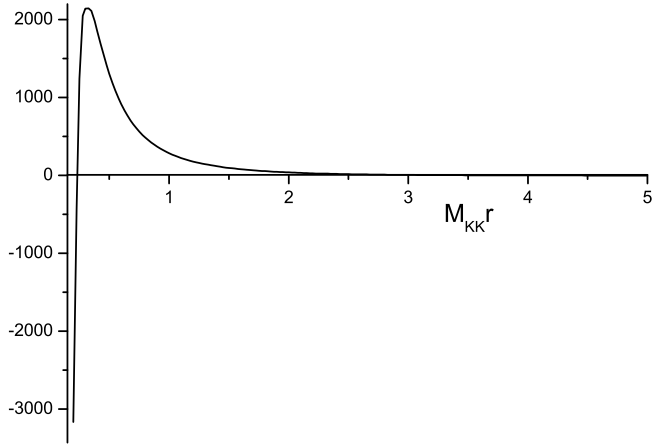}}
\par
\vskip-2.0cm{}
\end{center}
\caption{\small
This plot shows the large $N_c$ potential $V_{(p)}$
at short distance where the naive large $\lambda N_c$ formula
combined with the truncation becomes untrustworthy. Without the
finite $\lambda$-correction, the truncated potential turns
attractive again at a short distance of order $rM_{KK}\sim 3/p$.
The figure is for $p=10$.} \label{NN10-1}
\end{figure}

\begin{figure}[t]
\begin{center}
\scalebox{1}[1]{\includegraphics{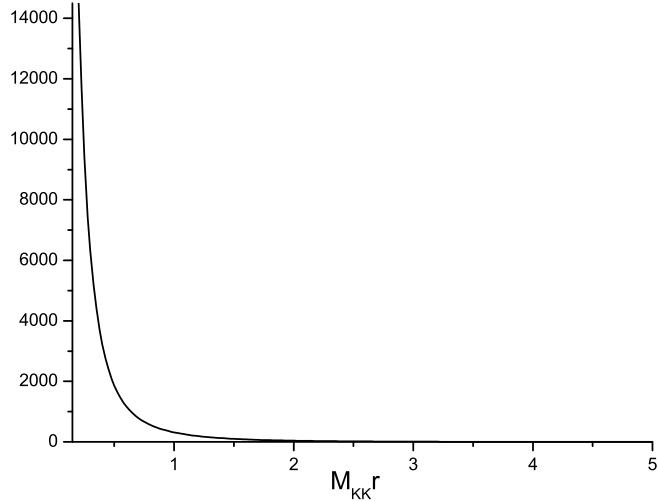}}
\par
\vskip-2.0cm{}
\end{center}
\caption{\small
This plot shows the large $N_c$ potential $V_{(p=10)}$
now with $1/\lambda$-corrected axial vector couplings at $\lambda=1100$.
The unphysical turnaround at $\sim 0.3M_{KK}$ disappeared
completely, allowing a smooth transition to the short distance
$1/r^2$ repulsive core. } \label{NN10-2}
\end{figure}

One reason behind this deficiency lies with the leading $\lambda N_c$
estimate we found in section 4. While most of estimate there are
safe in the large $N_c$ limit, the quantities $C_n^\pm$ are actually
correct only up to $\sim 1/\lambda$ corrections. This comes about because
\begin{equation}
\frac{2\pi^2\rho_{baryon}^2}{3e(w)^2}\simeq
\frac{\rho_{baryon}^2}{6}\times m_{\cal B}(w)\times
\left(1-\frac{\sqrt{2\cdot 3^5\cdot \pi^2/5}}{\lambda}+O(\lambda^{-2})\right)\,,
\end{equation}
implying that our numbers for $g_{a^{(k)}\cal NN}$ were
overestimated and we must adjust
\begin{equation}\label{shift}
 g_{a^{(k)}\cal NN}\quad\rightarrow\quad g_{a^{(k)}\cal NN}\times
 \left(1-\frac{\sqrt{2\cdot 3^5\cdot \pi^2/5}+}{\lambda}+O(\lambda^{-2})\right),
\end{equation}
if we wish to understand finite $\lambda$ cases, regardless of
$N_c\gg 1$.

Let us note that the smallest
distance for which we can trust the truncation up to the $p$-th pair
is around $rM_{KK}\sim 3/p$.
Comparing this distance against the solitonic baryon size, below which the
effective theory we used does not make much sense to begin with, we find that
such a truncated potential should be a sensible approximation
if we choose $p\sim \sqrt{\lambda/10}$.
For larger $p$, the idea of
point-like nucleon fails as far as interaction with heavier mesons are
concerned, while for smaller $p$ the potential $V_{(p)}$
fails at distances far larger than the individual baryon size.

For $\lambda\simeq 1000$, one may thus hope that the choice $p=10$
should be an optimal one. As we saw above, however,
the naive formula for the potential exhibits that
the potential begins to fails dramatically at $rM_{KK}=3/p$,
by turning strongly attractive again. Although there is no
strong inconsistency with this (since the potential failed
where it is expected to fail), it looks a little suspicious.
The point is that although $\lambda\simeq 1000$ seems large,
the correction (\ref{shift}) to
the axial vector couplings represents roughly
more than 2\% reduction and cannot be neglected.
What one should do is to correct $g_{a^{(k)}\cal NN}$ as in (\ref{shift})
and reconstruct the potential. Indeed, the numerical estimate shows
an almost complete cancellation of short distance $1/r^3$
when we take $p=10$ for the case of $\lambda=1100$. Figure 3
shows the corrected potential in this case, where the unphysical
turnaround at $rM_{KK}\sim 3/10$ disappeared. In the
case of $\lambda\sim 1000$, at least, the minimal choice for
truncation, $p=10$, was also effective.
Once this leaves behind $1/r^2$ terms as the leading
short-distance behaviors, the appearance of $1/r^2$ from the latter
via KK mode sum, where the vectors and the axial vectors
contributes with the alternating sign, follows easily.

For larger $\lambda$, the reduction of $g_{a^{(k)}\cal NN}$
would be smaller, but at the same time we must sum over more
mesons in order to make the potential
trustworthy down to the distance comparable to the soliton core size.
One may speculate that the optimal choice is again to sum up to
$p\sim \sqrt{\lambda/10}$. However, the couplings for
large $k$ are inherently ambiguous since it depends more and
more sensitively to, for example, the precise functional form of
the coefficient of $\bar {\cal B}{\cal F}{\cal B}$. This is because
the necessary mesonic wavefunction $\psi_{(n)}$'s are more and
more widespread, which also makes the couplings prone to
systematic errors from how the numerical estimate is cut-off
far away from  $w=0$.

%This also shows where and how to glue the
%intermediate and long distance potential we computed to the
%very short distance repulsive core consistently. Given finite
%$\lambda$, we should find an optimal $p$ for which the $1/r^3$
%short distance divergence is most effectively cancelled, and
%then cut-off the long distance form of the potential down to
%$r$ where the size of the remaining $1/r^3$ pieces can be at most
%comparable to that of $1/r^2$ pieces.

\section{Nucleon-Nucleon Potential for Realistic $\lambda$ and $ N_c$}

If one wishes to understand real QCD with $N_c=3$, one must
consider a different regime. For instance, we neglected $V_{\eta'}$ on account
of the small ratio
\begin{equation}
\frac{g_{\eta'\cal NN}}{g_{\pi\cal NN}}\sim\sqrt{\frac{3^7\cdot 5 \cdot\pi^2}{2^3}}\frac{1}{\lambda N_c}
\ll 1 \quad\hbox{when $\lambda N_c\gg 1$}
\end{equation}
in the holographic limit.
Yet, if we consider $N_c=3$ and $\lambda\simeq 17$ (determined
by measured values of $f_\pi$ ), we find
the ratio to be about 2 and is hardly ignorable. The estimates
here themselves are no longer reliable since we used large
$\lambda N_c$ limit, but this comparison clearly shows us that
we cannot expect any small parameter.
This is in fact a generic problem in going to the
realistic limit.

In computing Feynman diagrams and extracting nonrelativistic
potential, another small parameter is ${\mathbf p}/m_{\cal N}$
where ${\mathbf p}$ is the spatial momentum of the meson being
exchanged. However, when translated to real space, this ratio
can show up either as $m/m_{\cal N}$ or as $1/rm_{\cal N}$, which
is problematic when the meson mass $m$ exceeds the nucleon mass.
Thus, contribution from exchange of heavy mesons cannot be included
reliably, forcing us to cut down to pions, $\eta'$, $\rho$, and
$\omega$. Thanks to the universal suppression $\sim e^{-mr}$ for
heavy meson processes, this is a good approximation as long as
we are interested in distances strictly larger that $1/M_{KK}$.

The relevant Nucleon-Nucleon potential is then
\begin{equation}
V=V_\pi+V_{\eta'}+ V_{\rho^{(1)}}+ V_{\omega^{(1)}},
\end{equation}
where individual term must be computed as a series
expansion of $m/m_{\cal N}$. Actually,
the exchange of vector mesons generates a correction to
the kinetic term as well, the two-body Hamiltonian for
a pair of nucleons contains the relative part of the
Hamiltonian,
\begin{equation}
H=-\left(\frac{1}{m_{\cal N}}+\Delta\right)\nabla^2 +V
\end{equation}
with
\begin{equation}
\Delta=\frac{3m_{(1)}}{16\pi}\left(g_{\omega^{(1)}\cal NN}^2
+g_{\rho^{(1)}\cal NN}^2\vec \tau_1\cdot\vec\tau_2\right)
 \left(\frac{m_{(1)}}{m_{\cal N}}\right)^2y_0(m_{{(1)}}r)\,,
\end{equation}
where $ m_{(1)}\equiv m_{\omega^{(1)}} = m_{\rho^{(1)}}$.
In the attractive channel, $\vec\tau_1\cdot\vec \tau_2=-3$,
as we will see later, $(g_{\omega^{(1)}\cal NN}/g_{\omega^{(1)}\cal NN})^2\simeq 14$,
so the effective reduced mass of this two body system becomes
smaller as the distance becomes small.

Let us turn to the potential.
$V_\pi$ was already given in Eq.~(\ref{pion0}), while
others can be inferred from Ref.~\cite{EW88}. The contribution
from the trace part is essentially the same as the massive pion
case (\ref{pionfull}) except the $SU(2)$ generators $\vec\tau_1\cdot\vec\tau_2$
are absent
\bear
&&V_{\eta'}=\frac{1}{4\pi}\left (\frac{g_{\eta' \CN\CN}}{2m_\CN} M_{KK}\right )^2
\frac{m_{\eta'}^2}{M_{KK}^2}
\frac{m_{\eta'}}{3} [y_0(m_{\eta'}r)\vec\sigma_1\cdot\vec\sigma_2
+ y_2(m_{\eta'}r)S_{12}]\,.
\eear
The mass of $\eta'$ is generated by the $U(1)$ axial anomaly, and
was computed by Sakai and Sugimoto,
\begin{equation}
m_{\eta'}=\frac{\lambda M_{KK}}{\sqrt{27\pi^2}}\sqrt{\frac{N_F}{N_c}}\;.
\end{equation}
$V_{\rho^{(1)}}$ is considerably
more involved than $V_{\rho^{(1)}}^{holographic}$:
\bear
V_{\rho^{(1)}}&=&\frac{m_{{(1)}}}{4\pi}\biggl\{ \biggl [g_{\rho^{(1)}\CN\CN}^2
\left( 1-\frac14\frac{m_{{(1)}}^2}{m_\CN^2}\right)
+g_{\rho^{(1)}\CN\CN}\left(\frac{\tilde{g}_{\rho^{(1)}\CN\CN}}{2m_\CN}M_{KK}\right )
\frac{m_{{(1)}}}{m_\CN}
\frac{m_{{(1)}}}{M_{KK}}\nonumber\\
&&~~~~~+\frac14\left(\frac{\tilde{g}_{\rho^{(1)}\CN\CN}}{2m_\CN}M_{KK} \right)^2
\left(\frac{m_{{(1)}}}{m_\CN}\right)^2\left(\frac{m_{{(1)}}}{M_{KK}}\right)^2
\biggr]y_0 (m_{{(1)}} r)\nonumber\\
&&+\frac13\frac{m_{{(1)}}^2}{M_{KK}^2}\biggl[ \left(\frac{M_{KK}}{2m_\CN}g_{\rho^{(1)}\CN\CN}
+ \frac{\tilde{g}_{\rho^{(1)}\CN\CN}}{2m_\CN}M_{KK} \right)^2\nonumber\\
&&~~~~~+\frac18\left(\frac{\tilde{g}_{\rho^{(1)}\CN\CN}}{2m_\CN}M_{KK}\right)^2
 \left(\frac{m_{{(1)}}}{M_{\cal N}}\right)^2 \biggr]
\biggl[ 2y_0(m_{{(k)}} r)\vec\sigma_1\cdot\vec \sigma_2 -y_2(m_{{(k)}} r) S_{12}(\hat {r}) \biggr]\nonumber\\
&&-\left(\frac{m_{{(1)}}}{m_\CN}\right)^2\biggl[\frac32g_{\rho^{(1)}\CN\CN}^2 \nonumber\\
&&~~~~~+2g_{\rho^{(1)}\CN\CN}\tilde{g}_{\rho^{(1)}\CN\CN}
+\frac32\left(\frac{\tilde{g}_{\rho^{(1)}\CN\CN}}{2m_\CN}M_{KK}\right)^2
\left(\frac{m_{{(1)}}}{M_{KK}}\right)^2\biggr]
 \frac{y_1(m_{{(1)}}r)}{m_{{(1)}} r}\vec L \cdot\vec S\nonumber\\
&& +  \left(\frac{m_{{(1)}}}{m_\CN}\right)^4\biggl[\frac{1}{16}g_{\rho^{(1)}\CN\CN}^2
\nonumber\\
&&~~~~~+\frac12g_{\rho^{(1)}\CN\CN}\tilde{g}_{\rho^{(1)}\CN\CN}
+\frac12\tilde{g}_{\rho^{(1)}\CN\CN}^2\biggr]
 \frac{y_2(m_{{(1)}}r)}{m_{{(1)}}^2 r^2} Q_{12}\biggr\}\times \vec\tau_1\cdot\vec\tau_2\,,
 \eear
with
\bear
&&\vec S=\frac{1}{2}(\vec\sigma_1+\vec\sigma_2)\,,\nonumber \\
&&Q_{12}=\frac{1}{2} \left((\vec\sigma_1\cdot\vec L)(\vec\sigma_2\cdot\vec L)
+(\vec\sigma_2\cdot\vec L)(\vec\sigma_1\cdot\vec L)\right),
\eear
and  the spatial angular momentum $\vec L$.
Finally, $V_{\omega^{(1)}}$ is essentially of the
same form as $V_{\rho^{(1)}}$, except that
$\tilde g_{\omega^{(1)}\cal NN}=0$ and
$\vec\tau_1\cdot\vec\tau_2$ is absent,
 \bear
  V_{\omega^{(1)}}&=&\frac{m_{{(1)}}}{4\pi}g_{\omega^{(1)}\CN\CN}^2
  \biggl\{  \left( 1-\frac14\frac{m_{{(1)}}^2}{m_\CN^2}\right)
y_0 (m_{{(1)}} r)\nonumber\\
&&+\frac{1}{12}\left(\frac{m_{{(1)}}}{m_\CN}\right)^2\biggl[ 2y_0(m_{{(1)}} r)
\vec\sigma_1\cdot\vec \sigma_2 -y_2(m_{{(1)}} r)S_{12}(\hat {r})\biggr]\nonumber \\
&&-\frac32\left(\frac{m_{{(1)}}}{m_\CN}\right)^2 \frac{y_1(m_{{(1)}}r)}{m_{{(1)}} r}\vec L \cdot\vec S+
\frac{1}{16}\left(\frac{m_{{(1)}}}{m_\CN}\right)^4  \frac{y_2(m_{{(1)}}r)}{ m_{{(1)}}^2r^2} Q_{12}\biggr\}\, .
\eear
These are the complete expressions up to the quartic order
in terms of spatial momenta of individual nucleons.

For $N_c=3 $ and $\lambda \simeq 17$, we found the following
numbers that determine the couplings here,
\begin{equation}
4C_0^+ \simeq 0.697,\;\;2A_0^+\simeq 0.136,\;\;A_1^+\simeq 5.93,\;\;
2B_1^+\simeq \frac{7.04}{M_{KK}},\;\;2C_1^+\simeq -1.22
\end{equation}
and
\begin{equation}
f_\pi\simeq 0.0975M_{KK},\quad m_{(1)}\equiv m_{\rho^{(1)}}=m_{\omega^{(1)}}\simeq 0.818M_{KK},
\quad m_{\eta'}\simeq 0.85M_{KK}.
\end{equation}
The mass $m_{\cal N}$ has an inherent ambiguity since it would
be additively renormalized by massive excitations around the
soliton. Our definition of the nucleon mass kept only one such
massive mode, namely the position along $w$-direction, and according
to this prescription, we find
\begin{equation}
m_{\cal N}\simeq 1.93M_{KK}.
\end{equation}
Unfortunately, the scale
of $M_{KK}$ that fits the physical nucleon mass is about $\sim 500\,$MeV, as opposed to
the one needed to fit the physical $\rho $ meson mass  at $\sim 940\,{\rm MeV}$. This
discrepancy between the mesonic and the baryonic scales was previously
observed both in the D4-D8 modelin a slightly different comparison~\cite{Hata:2007mb}
and also in the so-called bottom-up approach \cite{Hong:2006ta}, and appears
unavoidable in the gravity approximation to the bulk side. We will proceed with
these numbers, nevertheless.
The couplings that enter the above potential are\footnote{It has been observed
previously that next subleading correction of some of the operators may
involve the simple shift $N_c \rightarrow N_c+2$ in the leading expressions.
This, for example, allows a very good match of $g_{\pi\CN\CN}$ with
experiment. The origin of this shift, originally suggested by the constituent
quark models, is not clear from this approach. Here, we chose not to implement
this shift but readers should be aware that terms from $\bar {\cal B}F{\cal B}$
may be affected, leading to quantitatively different numbers.}
\begin{eqnarray}
\frac{g_{\pi\CN\CN}}{2m_\CN} M_{KK} &=& \frac{4C_0^++2A_0^+}{2f_\pi}M_{KK}
\simeq 4.27 ,
\nonumber \\
\frac{g_{\eta' \CN\CN}}{2m_\CN} M_{KK} &=& \frac{2A_0^+\cdot N_c }{2f_\pi}M_{KK}
\simeq 4.18 ,
\nonumber \\
g_{\rho^{(1)}\CN\CN} &=& \frac{A_1^++2C_1^+}{2}
\simeq 2.36 ,
\nonumber \\
g_{\omega^{(1)}\CN\CN} &=& \frac{A_1^+\cdot N_c }{2}
\simeq 8.90 ,
\nonumber \\
\frac{\tilde{g}_{\rho^{(1)}\CN\CN}}{2m_\CN}M_{KK} &=&
\frac{2B_1^+\cdot M_{KK}}{2} \simeq 7.04 .
\end{eqnarray}
Detailed study of this case will be reported elsewhere.

\begin{figure}[t]
\begin{center}
\scalebox{1}[1]{\includegraphics{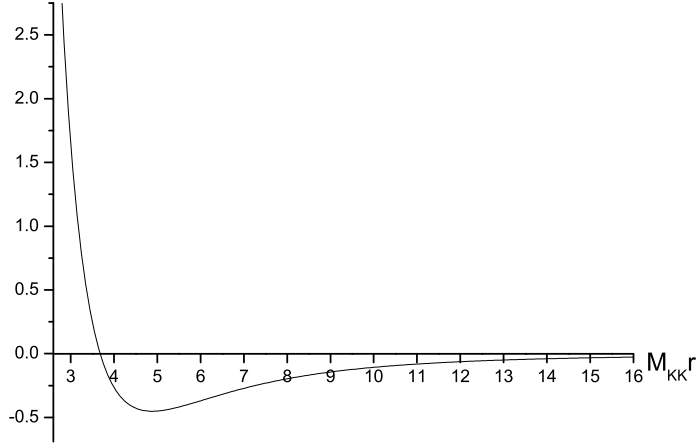}}
\par
\vskip-2.0cm{}
\end{center}
\caption{\small
A plot of the nucleon-nucleon potential, with $N_c\lambda=50$ and $N_c=3$.
We again drew the potential for the isospin singlet, $\sigma$-spin triplet,
and $S_{12}=2$ eigensector, although for this finite $\lambda N_c$
case we should expect different $S_{12}$ eigensectors to mix in.
This plot is only for the purpose of illustrating the general trend.
The horizontal axis is $rM_{KK}$, while the vertical potential
energy is in unit of $M_{KK}/4\pi$. } \label{NN1}
\end{figure}

\section{Holographic Deuteron: Large $N_c$ Results }

In this final section, we explore some basic aspects
of deuteron physics with the NN potential
\begin{equation}
V_\pi^{holographic}
+\sum_{k=1}^{10}\left( V_{\rho^{(k)}}^{holographic}+ V_{\omega^{(k)}}^{holographic}
+V_{a^{(k)}}^{holographic}\right)
\end{equation}
in the large $\lambda$ and $N_c$ limit. To distinguish this from physical
deuteron, we refer to them as holographic deuterons.

For a bound state, we need  to focus on the long distance
attractive channel. The large $N_c$ potential has a simple spin and
flavor structure as
\begin{equation}
V^{holographic}=V_C+(V_T^{\sigma}\vec\sigma_1\cdot\vec\sigma_2+
V_T^{S} S_{12})\,\vec\tau_1\cdot\vec\tau_2 .
\end{equation}
The massless pion exchange, the dominant contribution
in long distance, contributes only to $V_T^{S}$ and positively,
so an attractive channel requires
$S_{12}\vec\tau_1\cdot\vec\tau_2 <0$. Using the fact that $S_{12}$
acting on $\sigma$-spin singlet vanishes identically, and
that the nucleons are fermions, this forces the isospin singlet
($\vec\tau_1\cdot\vec\tau_2=-3$) and the $\sigma-$triplet
($\vec\sigma_1\cdot\vec\sigma_2=1$) channel with even spatial
angular momentum. The lowest total angular momentum possible is
then $J=1$, and the positive $S_{12}$ eigensector has the following
spatial angular momentum mix as
\begin{equation}
\frac{|L=1\rangle+\sqrt{2}|L=0\rangle}{\sqrt{3}}.
\end{equation}
In this eigensector, $S_{12}=2$.
The figure 1 is the plot of the potential in this sector,
whose classical minimum occurs at
\begin{equation}
r_{min}\simeq \frac{5.53}{M_{KK}}\;, \qquad V(r_{min})\simeq-0.0944M_{KK}{\frac{N_c}{\lambda}}.
\end{equation}
Note that the binding potential is very shallow.  Recall that
in the large $\lambda N_c$ limit, $M_{KK}$ is most conveniently
determined by the  vector meson scale to be around $0.94$~GeV. Among
various scales that enter the baryon energy, we have the
hierarchies,
\begin{equation}
m_{\cal N}\sim m_{B}^{classical}\sim \lambda N_c \quad\gg\quad E_{Coulomb}\sim N_c
\quad\gg\quad
|V(r_{min})|\sim {\frac{N_c}{\lambda}}.
\end{equation}
The middle measures the energy related to the classical deformation
of the individual baryon away from the self-dual  soliton,
while the last measures the binding energy of the nuclei.

The fact that the nuclei biding energy is small, which is also
borne out in real world,\footnote{The physical deuteron has a binding
energy of $2.2\,$MeV \cite{Garcon:2001sz}, which is about 0.12\%
of its rest mass.} is interesting from the
standpoint of the holographic QCD as well. The stringy
picture of the baryon says that the individual baryon can
be viewed as a D4 brane wrapped on the compact $S^4$ of the
dual geometry \cite{witten:1998xy,sakai-sugimoto}. What we computed here is essentially the
potential between two such objects separated along the
noncompact $R^3$. The binding energy is positive but
suppressed relative to the individual rest mass by
$1/\lambda^2$, indicating very weak interactions. In terms of the warped
string scale, $\alpha'_{warped}$, this power is
equals $(\alpha'_{warped}M_{KK}^2)^2$.
Although the significance of this particular power is
unclear to us, it does show that the two wrapped D4
branes are almost non-interacting at long distances.
This seems to suggest that the object underlying baryons
may remain close to its original BPS nature, despite the
supersymmetry breaking background of
scale $M_{KK}$ and high mass $\sim \lambda N_c M_{KK}$,
which is well beyond the cut-off scale $M_{KK}$, and may
eventually
explain why such a high mass object is well-described
by this D4-D8 holographic QCD.

\section{Concluding Remarks}

In this work, we computed the nucleon-nucleon potential
in the D4-D8 holographic QCD, which is generated by exchange
of five-dimensional flavor gauge field. In four dimensional
picture, this amounts to exchange of massless pseudo-scalars
and an infinite tower of spin one mesons. In the large $\lambda
N_c$ limit, it is sensible to sum up to first $\sim \sqrt{\lambda/10}$
vector and axial vector meson pairs, although one may choose to cut it shorter
according to the shortest distance scale interested. This prescription
also gives whereto glue the repulsive short distance regime to
the more complicated intermediate and long distance regime.
Some rudimentary
aspects of deuteronic bound state is explored for large $N_c$
case. Consideration of deuteron for realistic QCD regime
will be explored elsewhere.

We hope this work will provide a more practical starting point
for exploration of how holographic QCD fares against experimental
data, part of which comes from nucleon-nucleon scattering
amplitudes. Admittedly, this would involved huge extrapolation
to $N_c=3$ and $\lambda\simeq 17$, where the holographic
approach is hardly justifiable by the first principle. But,
in the absence of any other honest derivation
of nucleon-nucleon potential, our result should be at least
tested against data. In this work, we did not attempt to analyze
realistic QCD regime and concentrated mostly on large $N_c$
limit. We wish to come back later to the $N_c=3$ potential of
section 6, and explore its consequences.

Another important application of this work
would be in understanding dense matter system, such as
neutron stars, where the correct equation of state is of
some importance. In dealing with such a dense system from
the holographic QCD, baryon density itself were often
treated as external input in the form of delta-function
density in five-dimensions. We hope our nucleon-nucleon
potential would allow a more refined approach.

\vskip 1cm

\subsection*{Acknowledgments}

P.Y. is grateful to Lenny Susskind for a comment that motivated
this work, and also Deog-Ki Hong, Shamit Kachru, Mannque Rho, and Ho-Ung Yee for
discussions. He also thanks SITP of Stanford University for hospitality
and generous support. Y.K. thanks Hyun-Chul Kim for useful comments.
Y.K. acknowledges the Max Planck Society(MPG) and the Korea Ministry of
Education, Science and Technology(MEST) for the support of the Independent
Junior Research Group at the Asia Pacific Center for Theoretical Physics
(APCTP).
S.L. is supported in part by the KOSEF Grant R01-2006-000-10965-0 and
the Korea Research Foundation Grant KRF-2007-331-C00073.
P.Y. is supported in part by the Science Research Center Program of KOSEF
(CQUeST, R11-2005-021), the Korea Research Foundation (KRF-2007-314-C00052),
and by the Stanford Institute for Theoretical Physics (SITP Quantum Gravity visitor fund).

\end{document}